\documentclass[a4paper,twocolumn,11pt,accepted=2025-06-30]{quantumarticle}
\pdfoutput=1
\usepackage[numbers]{natbib}

\usepackage{graphicx}
\usepackage{dcolumn}
\usepackage{bm}
\usepackage{algorithm}
\usepackage{algpseudocode}
\usepackage{float}
\usepackage{braket}
\usepackage{lipsum}
\usepackage{orcidlink}
\usepackage[hashEnumerators,smartEllipses]{markdown}

\usepackage[caption=false,listofformat=subsimple,labelformat=simple]{subfig}

\usepackage[export]{adjustbox}

\usepackage{textcomp} 

\usepackage{tabu}

\usepackage{color}
\definecolor{darkred}{RGB}{150,0,0}

\usepackage{xparse}
\RenewDocumentCommand\log{g}{%
  \IfNoValueF{#1}{\text{log}\left(#1\right)}%
  \IfValueF{#1}{\text{log}}%
}
\usepackage{hyperref}
\newcommand{\etal}{\textit{et~al.} }

\newcommand{\brakets}[2]{\langle #1|#2 \rangle}
\newcommand{\ketbra}[2]{| #1 \rangle\!\langle #2 |}
\newcommand{\ketbraS}[3]{| #1 \rangle_{#3}\!\langle #2 |}

\DeclareMathOperator{\Tr}{Tr}
\begin{document}


\title{%
Learning Feedback Mechanisms for Measurement-Based Variational Quantum State Preparation
}%

\author{Daniel Alcalde Puente \orcidlink{0000-0002-3519-5931}} 
\affiliation{Forschungszentrum Jülich, Institute of Quantum Control, Peter Grünberg Institut (PGI-8), 52425 Jülich, Germany}
\affiliation{Institute for Theoretical Physics, University of Cologne, 50937 Köln, Germany}




\author{Matteo Rizzi \orcidlink{0000-0002-8283-1005}}
\affiliation{Forschungszentrum Jülich, Institute of Quantum Control,
Peter Grünberg Institut (PGI-8), 52425 Jülich, Germany}
\affiliation{Institute for Theoretical Physics, University of Cologne, 50937 Köln, Germany}

\begin{abstract}
This work introduces a self-learning protocol that incorporates measurement and feedback into variational quantum circuits for efficient quantum state preparation. By combining projective measurements with conditional feedback, the protocol learns state preparation strategies that extend beyond unitary-only methods, leveraging measurement-based shortcuts to reduce circuit depth.
Using the spin-1 Affleck-Kennedy-Lieb-Tasaki state as a benchmark, the protocol learns high-fidelity state preparation by overcoming a family of measurement-induced local minima through adjustments of parameter update frequencies and ancilla regularization. Despite these efforts, optimization remains challenging due to the strongly non-convex optimization landscapes inherent to variational circuits.
The approach is extended to larger systems using translationally invariant ansätze and recurrent neural networks for feedback, demonstrating scalability. Additionally, the successful preparation of a specific AKLT state with desired edge modes highlights the potential to discover new state preparation protocols where none currently exist.
These results indicate that integrating measurement and feedback into variational quantum algorithms provides a promising framework for quantum state preparation.
\end{abstract}

\maketitle
\section{Introduction}

Quantum technologies have significant potential to address key challenges in quantum simulation, communication, and information processing. As such, the efficient preparation of high-fidelity quantum states and the creation of robust state preparation protocols are critical, particularly for noisy intermediate-scale quantum (NISQ) devices~\cite{Preskill2018quantumcomputingin}. Additionally, quantum measurements are becoming an integral part of many new circuit paradigms~\cite{fisher2023random,zhu2023nishimori, bravyi2022adaptive,sukeno2024quantum}, emphasizing the importance of incorporating measurement and feedback mechanisms into quantum state preparation. In this work, we explore the autonomous learning of variational quantum circuits with measurement and feedback for quantum state preparation.

Variational quantum circuits (VQCs)~\cite{cerezo2021variational} hold promise for a wide range of quantum computing applications.
A parameterized quantum circuit is constructed from gates parameterized by their angles. 
Its parameters are optimized to minimize a cost function, often infidelity, in the context of state preparation. 
Preparing one-dimensional long-range entangled quantum states using local two-qubit variational quantum circuits (VQCs) typically requires a circuit depth that scales linearly with the system size.

However, measurement-based quantum circuits can be analytically derived for constant-depth state preparation by exploiting the non-unitary character of quantum measurements and the use of adaptive feedback. These methods have been applied to a broad class of target states, including toric code states, symmetry-protected topological (SPT) phases, fracton phases, and Schrödinger’s cat states~\cite{Tantivasadakarn2024SPTMeasurement, Lu2022MeasurementShortcut, verresen2021efficiently, piroli2021quantum}, as well as certain non-Abelian topological orders~\cite{Tantivasadakarn2023non-abelian}. Adaptive measurement strategies further extend these capabilities by enabling the fusion of small resource states into complex many-body targets. This approach has led to deterministic, constant-depth protocols that have been demonstrated experimentally, for instance, in the preparation of the AKLT state~\cite{smith2023deterministic}. The feasibility of such fusion-based schemes depends on the entanglement structure of the target state. Theoretical results identify features such as flat entanglement spectra in SPT phases as crucial for exact preparation~\cite{stephen2024preparing,lootens2023low,sahay2024classifying}, and also extend to symmetry-broken and topologically ordered phases~\cite{smith2024constant}. Allowing for approximate rather than exact preparation further reduces circuit complexity. For example, translationally invariant matrix product states can be prepared approximately with depth scaling as  $\mathcal{O}(\log\log(N/\epsilon))$~\cite{malz2024preparation}, where $\epsilon$ is the allowed error. In addition, probabilistic schemes permit constant-depth preparation of states such as Dicke and W-states by accepting non-deterministic outcomes~\cite{piroli2024approximating}.

Incorporating these ideas, this work extends VQCs to include non-unitary measurements and feedback, enabling the protocol to learn fundamentally different state preparation techniques than standard VQCs and potentially reducing the required circuit depth. 
These measurement-based protocols exploit the non-unitary nature of quantum measurements to construct non-unitary gates. 
However, one cannot apply non-unitary gates directly; rather, one must design a set of non-unitary gates that, when combined, form a completely positive trace-preserving (CPTP) map from which one gate is randomly applied. 
The challenge is to design a CPTP map that, through the application of conditional feedback, ensures that the desired quantum state is achieved regardless of the specific non-unitary gate that is randomly selected by quantum mechanics. 
In this work, this task is learned automatically using VQCs.

Despite their potential, VQCs face significant challenges that hinder their practical implementation. One prominent problem is the occurrence of barren plateaus in the optimization landscape of deep circuits, where the gradients of the parameters approach zero, rendering the optimization process infeasible. 
While measurement-based methods at shorter circuit depths may avoid these plateaus, they introduce new optimization challenges. 
Shallow quantum circuits, while less prone to barren plateaus, can be difficult to optimize due to the presence of local minima~\cite{anschuetz2022swamped, cerezo2021shallow}. Local minima can be partially mitigated in specific cases, such as the AKLT model, by incorporating symmetry constraints into the variational ansatz~\cite{lyu2024variational}. However, there is no general strategy for addressing this issue across different models.
In this work, additional local minima that are unique to measurement-based VQCs are discussed and mitigation strategies are discussed.

Autonomous learning of measurement and feedback protocols for state preparation have been studied in the literature with two different methods.

In the first method, reinforcement learning is used to develop a complete policy that determines dynamically both when measurements should be performed and which unitary operations should be applied based on those measurements.
This approach shows success in the single-particle case~\cite{Wang2020Cartpole, Borah2021DoubleWell} and for two-particle systems~\cite{Sivak2022Model-Free, puviani2023boosting}. However, we note that given the evidence presented in the papers we expect these methods to have significant challenges when extended to multi-qubit systems.

In the second method, greedy optimization techniques have been proposed~\cite{Gefen2023Active, Morales2024unstearable} to prepare multi-qubit states.
These methods periodically optimize unitaries to maximize fidelity after each weak measurement.
More precisely some ancillas $A$ are introduced and coupled to the system with unitaries and the measured projectively. 
Each of these unitaries is learned independently like:
$\ket{\psi_i'(\theta_i,...)}=U(\theta_i)\ket{\psi_{i-1}(\theta_{i-1}, ...)}$ with $\theta_i =\min_{\theta_i} \text{ loss}(\ket{\psi_i'(\theta_i,...)}$,
and $\ket{\psi_i(\theta_i,...)}  = \bra{M}_A\ket{\psi_i'(\theta_i,...)}$ sampled from $M \in \lVert\bra{M}_A \ket{\psi_i'(\theta_i,...)}\rVert^2$.
This makes them suffer from two major shortcomings.
First, implementing these methods experimentally requires running a simulated version of the experiment in parallel to optimize the unitary operations on the fly depending on the measurement results obtained in the experiment.
Second, the cost function used in these approaches is inherently greedy, reacting to measurement results rather than proactively incorporating them into the optimization. Consequently, these limitations hinder the preparation of states with more than a few qubits and prevent the learning of feedback mechanisms such as those proposed by Smith~\etal~\cite{smith2023deterministic}, where the pre-measurement unitary circuit is designed in such a way that after the measurement all possible resulting states can be mapped to the target state using two-body gates. For this, all unitaries must be learned at the same time for all possible measurement outcomes, which is done in our work.

Our learning technique can handle larger systems than the aforementioned protocols using reinforcement learning because it has a fixed structure (VQCs), simplifying the learning problem. 
However, since our protocol is non-greedy and learns the complete feedback step  by learning all unitaries at the same time, it can effectively learn which CPTP map and corresponding feedback to apply automatically, and as such can learn more intricate protocols like the one by Smith~\etal~\cite{smith2023deterministic}.

Mixing variational circuits and measurement has also been discussed in the realm of Measurement-based quantum computing (MBQC), where VQE can be used to decide which basis to measure with during the circuit~\cite{chan2024measurement}.

Similar concepts have been explored in the study of error-correcting codes, where reinforcement learning algorithms attempt to automate the discovery and implementation of error-correcting protocols. 
The complexity of this task forces its decomposition into subtasks that are handled by different reinforcement learning agents, such as the decoding of known stabilizer codes~\cite{Eisert2020Decoders, Andreasson2019quantumerror, Laia2020RLerror}, or the spatial deformation of codes for better logical error rates~\cite{Nautrup2019optimizingquantum}. 
Although learning the complete error-correcting task simultaneously would be ideal, it remains highly challenging and is only accomplished with strategies that do not scale with system size in~\cite{Marquardt2018Reinforcement} for a small system of four qubits.
This highlights the difficulty of learning full measurement and feedback-based protocols, which our work addresses in the context of state preparation.

\label{sec:alam_aknowledgement}
During the final preparation of this manuscript, various manuscripts were published using a similar framework. In \cite{ilin2024dissipative}, a similar approach is employed but without measurement feedback, relying solely on projective measurements as a non-unitary operation. In \cite{yan2024variational, alam2024learning}, the protocols largely mirror those presented here, with minor variations, but applied to the GHZ state. In Yan~\etal~\cite{yan2024variational}, the authors propose a method to experimentally estimate the gradients of the protocol’s parameters on a Quantum Computer. In Alam~\etal~\cite{alam2024learning}, a Density Matrix Renormalization Group (DMRG)-inspired sweeping optimization is introduced to avoid measurement-induced local minima in the optimization landscape. For completeness, we have added a short analysis of the GHZ state using our methodology in \ref{appendix:mvqe:ghz}.

However, we expect this to worsen the above-mentioned vanilla-VQC local minima~\cite{anschuetz2022swamped, cerezo2021shallow}, which should still be present in addition to measurement-induced ones. Our internal experiments on vanilla-VQCs (inspired by Pollmann \etal~\cite{pollmann2016efficient}) indicate indeed that sweeping does not scale well with system size and tends to get trapped in local minima for larger systems. Therefore, the application of a pure sweeping approach to larger measurement-based VQCs remains unclear to us and constitutes an interesting direction for future research. Consequently, we did not pursue this approach in our work, although a hybrid approach could be beneficial. Instead, we found an explanation for why the measurement-induced local minima occur and addressed them directly for a system size of 16 qubits.

This paper is organized as follows: 
In Sec.~\ref{sec:mvqe:ProtocolDescription}, we present our self-learning protocol that integrates measurement and feedback into VQCs for efficient quantum state preparation. 
The protocol employs a sequence of parameterized unitaries and projective measurements, using feedback from measurement outcomes to inform subsequent operations. 
This approach allows the protocol to learn non-unitary state preparation techniques that can reduce the circuit depth.
In Sec.~\ref{sec:mvqe:AKLT} and ~\ref{sec:mvqe:local_minima}, we apply our protocol to the preparation of the spin-1 AKLT state, using it as a benchmark to evaluate the learning capabilities of our approach, which if optimized naively is plagued by local minima. 
We propose two strategies to mitigate these: adjusting the parameter update frequencies between the initial unitary and feedback operations (detailed in Sec.~\ref{sec:mvqe:different_update_frequencies}), and introducing ancilla regularization to promote a more uniform distribution of measurement outcomes (discussed in Sec.~\ref{sec:mvqe:ancilla_regu}).
In Sec.~\ref{sec:mvqe:comparison}, we compare the learned protocol to the analytically derived protocol proposed by Smith~\etal~\cite{smith2023deterministic}.
We analyze the similarities and differences using quantum mutual information, highlighting how the learned protocol can achieve similar or better performance with potentially shallower circuits by requiring less mutual information prior to measurement.
In Sec.~\ref{sec:mvqe:PeriodicAnsatzRNN}, we extend our approach to larger systems by employing a translationally invariant ansatz and utilizing a  Recurrent Neural Network (RNN)  for the feedback function. 
We demonstrate our protocol's scalability and discuss the RNN's performance, but note challenges in performing optimal corrections for large systems.
In Sec.~\ref{sec:mvqe:SpecificAKLT}, we explore the preparation of a specific AKLT state with both edge modes in the spin-up configuration—a task for which no known deterministic, low-depth protocol exists. 
We show that our learning protocol can discover such a state preparation strategy, highlighting its potential to find new protocols where none currently exist.
In conclusion, this study shows the promise of integrating measurement and feedback into variational quantum algorithms to discover quantum state preparation algorithms.


\section{Protocol Description}
\label{sec:mvqe:ProtocolDescription}
\begin{figure*}[t]
    \centering
    \subfloat[]{\raisebox{0.1\height}{\label{fig:mvqe:ansatz:hardware_efficient}\includegraphics[width=0.49\textwidth]{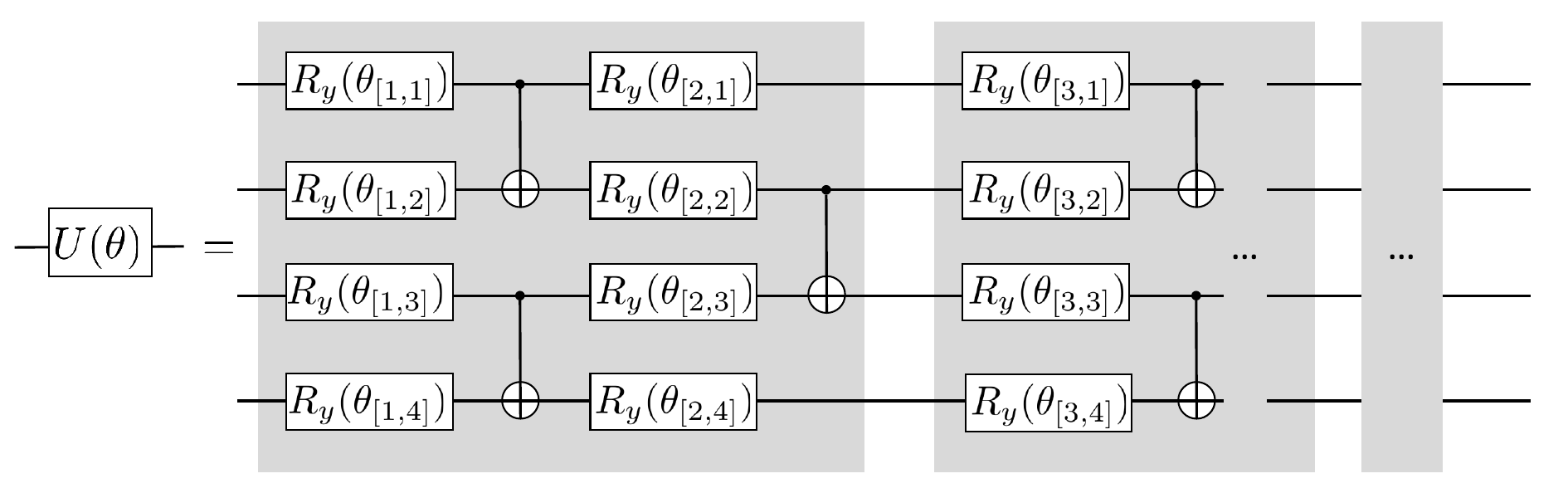}}}
    \hfill
    \subfloat[]{\label{fig:mvqe:ansatz:protocol}\includegraphics[width=0.49\textwidth]{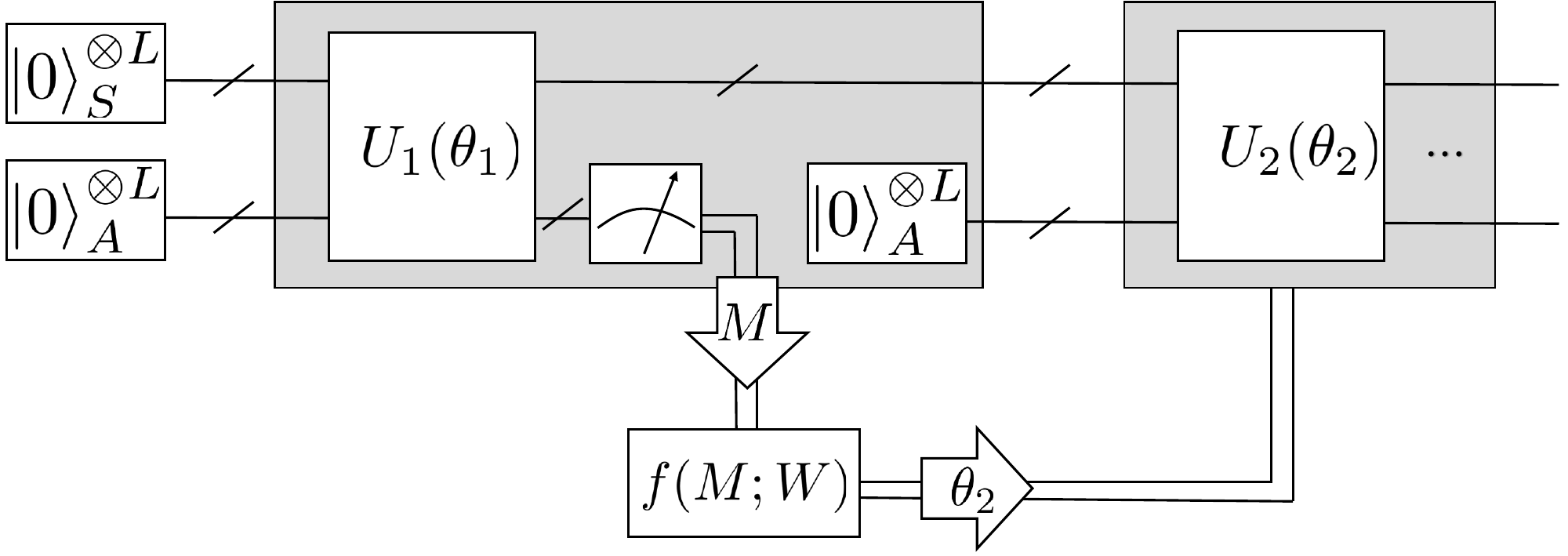} }
    
    \caption{Illustration of the quantum feedback control protocol.
    (a) Depicts the hardware-efficient ansatz used in the construction of the unitary $ U(\theta) $, showcasing a series of parameterized rotation gates $ R_y(\theta_{i,j}) $ arranged in alternating layers with CNOT gates.
    (b) Outlines the quantum-classical feedback loop, starting with the initial state preparation of the system $S$ and ancilla $A$ in $|0\rangle$ states. The application of $U(\theta_1)$ is followed by measurement, and the measurement results $M$ are fed into a function that outputs the parameters $\theta_2=f(M;W)$ for the next unitary operation $U(\theta_2)$. This loop implements the adaptive adjustment of parameters based on measurement results, which is central to the feedback control strategy. Note that the optimal parameters $\{\theta_1, W\}$ are learned using the gradient descent algorithm.}
    \label{fig:mvqe:ansatz}
\end{figure*}%

Our protocol consists of a sequence of parameterized unitary transformations and projective quantum measurements to prepare a desired target state $\ket{\psi_t}$. 
This process is iterative, using feedback from measurement outcomes to inform subsequent quantum operations. 
Its simplest form, with only one feedback round in the ansatz, is reminiscent of error correction and can represent state preparation protocols like~\cite{Tantivasadakarn2024SPTMeasurement, Tantivasadakarn2023non-abelian, Lu2022MeasurementShortcut, smith2023deterministic, piroli2024approximating, smith2024constant, stephen2024preparing, sahay2024classifying}.
Conversely, if the feedback is removed completely (set it to a constant), then the operation $U(\theta)$ with the addition of measurement and reset, will be able to represent passive steering strategies like~\cite{alcalde2024quantum, roy2020measurement, matthies2022programmable}. 
This emphasizes the representative capacity of this approach. 
The potential of this generalized, learnable protocol is twofold: first, to discover methods for preparing quantum states where no effective protocol currently is known; and second, to identify faster and more optimal approaches for state preparation where established protocols are already known.


\subsection{Framework}
Our protocol, shown in Fig.~\ref{fig:mvqe:ansatz:protocol}, is described by the following sequence of operations:
\begin{align}
\rho_1(\theta_1) &= U_{1}(\theta_1) \rho_0 U_{1}^\dagger(\theta_1), \\
\rho_{1}^M(\theta_1) &=  \ketbraS{0}{M}{A} \rho_1(\theta_1) \ketbraS{M}{0}{A}, \label{eq:mvqe:rhoM}\\
\rho_2(\theta_1,W) &= \sum_M U_{2} (\theta_{2}) \rho_{1}^M(\theta_1) U_2^\dagger(\theta_2), \\
\text{with} & \ \ \theta_2=f(M;W) \nonumber
\end{align}
where $\rho_0$ is the initial state of the system, which we will choose to be $\rho_0=\ketbraS{0}{0}{SA}$, $U(\theta)$ is a unitary operation parameterized by $\theta$, and $M$ are the measurement outcomes.
The operator $\ketbraS{0}{M}{A}$ projects the ancilla qubits to the measured state $M$ which is then reset to a product state composed of zeros.
The feedback function $f(M;W)$ with learnable parameters $W$ is the key ingredient of the protocol as it decides on the parameters of subsequent unitary operations based on previous measurement outcomes.

For a more generalized approach that can be applied iteratively for multiple rounds of feedback, we extend the notation as follows:
\begin{align}
\rho^M_i(\theta_1,W_2,...,W_i) & = \ketbraS{0}{M}{A} \rho^{ }_i \ketbraS{M}{0}{A} \\
\rho^{ }_i(\theta_1,W_2,...,W_i) & = \sum_M U_i(\theta_i) \rho^{M}_{i-1} U^\dagger_i(\theta_i) \\
\text{with} \ \ \theta_i & = f_i( M; W_i) \nonumber
\end{align}
This iterative relation frames any general feedback protocol within our proposed scheme, where different functions $f_i(M; W_i)$ decide on subsequent unitary operations. Depending on the ansatz it is possible to use different parameterized unitaries $U_i$ for different feedback sweeps.

The efficacy and applicability of the proposed protocol will first be evaluated in the context of the AKLT state preparation, demonstrating its ability to learn feedback strategies. Specifically, we aim to automatically reproduce the deterministic low-depth protocol outlined in~\cite{smith2023deterministic}, which will be further discussed in Sec.~\ref{sec:mvqe:AKLT}.


\subsection{Feedback Mechanism}

The success of our protocol is critically dependent on the function $f(M;W)$, which parametrizes the feedback loop. 
It maps measurement outcomes onto new parameters for unitary transformations. 
This function is vital, as it chooses subsequent quantum gates depending on the measurement results.

For small ancilla qubit Hilbert spaces, $f(M;W)$ can be effectively represented using a tabular approach. 
In this representation, $f(M;W) = W_M$, with $W_M$ being a set of learnable vectors, each of them storing the angles for the parameterized unitaries that should be applied in response to a particular measurement outcome $M$.
This tabular method is straightforward and computationally manageable when the ancilla Hilbert space remains small.

However, as the ancilla Hilbert space expands, the tabular method becomes impractical because of the exponential increase in the number of potential measurement outcomes. 
For larger systems, a neural network offers a more sophisticated and scalable representation for $f(M;W)$. 
The chosen architecture for our implementation includes both SwiGLU (Swish-Gated Linear Unit) and bidirectional Recurrent Neural Network (RNN) layers, which were chosen as they are adept at modeling spatial correlations. 
These are used in Sec.~\ref{sec:mvqe:PeriodicAnsatzRNN} to obtain a translationally invariant feedback ansatz and their structure is explained in detail in App.~\ref{appendix:mvqe:rnn}.


\section{AKLT State(s)}
\label{sec:mvqe:AKLT}

The 1D spin-1 Affleck-Kennedy-Lieb-Tasaki (AKLT) states~\cite{Affleck1987AKLT} provide a useful benchmark for evaluating the learning capabilities of our protocol. These states are well-known for their symmetry-protected topological (SPT) properties, characterized by two spin-$\frac{1}{2}$ edge states~\cite{verresen2017one, pollmann2012detection}. 

The presence of these two free spin-$\frac{1}{2}$ edge states results in a four-fold degeneracy of the AKLT state. Furthermore, the AKLT state has been proposed as a valuable resource for measurement-based quantum computing~\cite{brennen2008measurement}.
If just local unitaries are employed, the preparation time scales linearly with the system size: 
by using also measurements, Smith~\etal have shown that a protocol with depth independent from the system size is possible~\cite{smith2023deterministic}.
This consists of preparing small AKLT states and fusing them using measurement and deterministic feedback based on the measurement outcomes. This method is deeply rooted in the exact representation of AKLT states as matrix product states.

The analytically derived protocol in~\cite{smith2023deterministic} can prepare a random state in the manifold spanned by the four AKLT states within a single round of measurement and feedback. Thus, our first task will be to try to learn an equivalent protocol without any prior knowledge of~\cite{smith2023deterministic}, to better understand the challenges faced by the learning algorithm.

Under the restriction of using a single round of measurements, the fidelity to the AKLT manifold, spanned by $\ket{\psi_i}$, can be computed via:
\begin{align}
\label{eq:mvqe:loss}%
    F(\theta_1,W) & = \sum_i^4 \bra{\psi_i}_S \rho_2(\theta_1,W) \ket{\psi_i}_S \\ 
    & = \sum_i^4\sum_M \lVert \bra{\psi_i}_S U_2^{(S)}(\theta_2) \bra{M}_A U_1^{(SA)}(\theta_1) \ket{0}_{SA} \rVert^2 \nonumber \\ 
    &\qquad \qquad \qquad
    \text{with}  \ \ \theta_2=f(M;W) 
\end{align}
where $U_1^{(SA)}(\theta)$ represents a hardware-efficient ansatz (see~Fig.~\ref{fig:mvqe:ansatz:hardware_efficient}), and $U_2^{(S)}(\theta)$ is a similar ansatz acting only on the system qubits.
Ideally, the feedback operation would use an ansatz identical to $U^1_{SA}$; however, implementing this approach with a short circuit depth after measurement introduces insurmountable local minima. 
This issue commonly arises in shallow VQCs when the initial wave function is not a product state, making optimization highly challenging. Therefore, instead of replicating $U_1^{(SA)}$, a simpler ansatz using only two-qubit gates is selected for the feedback step. Further details are provided in App.~\ref{app:hardware_efficeint_two}.

For the sake of simplicity, we have first restricted ourselves to just employing tabular feedback $f(M;W) = W_{M}$ as it offers fewer possible points of failure than the use of a Neural Network, which we later analyze in Sec.~\ref{sec:mvqe:PeriodicAnsatzRNN}.
The objective is to optimize this protocol by minimizing the infidelity. 
We employ a gradient descent approach, specifically using the ADAM optimizer~\cite{kingma2014adam}, for this purpose. 
However, a straightforward application of this optimization method leads to entrapment in local minima, as do other commonly used optimization methods like Natural Gradient, LBFGS, and gradient-free methods.
These local minima are distinct both from the barren plateaus encountered in long and wide circuits and from local minima encountered in shallow circuits~\cite{anschuetz2022swamped, cerezo2021shallow}. 
To circumvent this class of local minima specific to feedback protocols, we implement two novel strategies as described in Sec.~\ref{sec:mvqe:different_update_frequencies} and Sec.~\ref{sec:mvqe:ancilla_regu}.

Note that in this work the spin-1 is mapped to qubits with the mapping $\ket{+}\rightarrow\ket{10}$, $\ket{0}\rightarrow\ket{00}$ and $\ket{-}\rightarrow\ket{01}$ as done also in~\cite{smith2023deterministic}, as it reduces entanglement between the two spin $\frac{1}{2}$, without affecting the properties of the final state. Other encoding choices are also possible and may impact the circuit's efficiency, as discussed in Ref.~\cite{sawaya2020resource}, where various mappings were analyzed in the context of Trotterization.

\section{Local minima and how to avoid them}
\label{sec:mvqe:local_minima}
During the optimization process, it was observed that the probability distribution $P(M;\theta) = \Tr_S\left[\langle M|_A \rho_i(\theta) |M\rangle_A\right]$ for measuring the ancillas in a particular bit-string $M$ in the z-base tends to favor a specific value of $M'$. 
This results in $P(M;\theta) = \delta(M,M')$ being a delta function, so that when the ancillas are measured, they always give the same measurement result $M'$. 
This results in the measurement operation having no effect on the quantum state, indicating that the optimization algorithm effectively found a way to bypass the measurement operation and as such the feedback altogether.

This can be quantified by the Shannon entropy $H$ or the harder-to-compute entanglement entropy $S$, defined as:
\begin{align}
    \rho_1^A &= \Tr_S(U_{1}(\theta_1) \ketbraS{0}{0}{SA} U_{1}^\dagger(\theta_1)) \\
    P(M) &= \bra{M} \rho_1^A \ket{M} \\
    H &= -\sum_M P(M)\log_2(P(M)) \\
    S &=-\Tr_A[\rho_1^A \log(\rho_1^A)]\,.
\end{align}
Both of them turn out to be small, or even zero, at the local minima encountered along a naive minimization of the loss function in Eq.~\ref{eq:mvqe:loss}, as can be seen in the orange curve in Fig.~\ref{fig:mvqe:freq_ancilla_regu=0:entropy}.
In the next two sections, two strategies to mitigate local minima are discussed.

\subsection{Different update frequencies}
\label{sec:mvqe:different_update_frequencies}

We conjecture that the extreme sharpening of the measurement probability distribution $P(M)$ occurs due to an imbalance between the intrinsic learning rates of the initial unitary $U_1(\theta_1)$ and of the feedback $U_2(\theta_2)$, with the latter being too slow.
The first attempt to improve the optimization involves therefore an ad-hoc increase of the update frequency of the $W$ parameters that define the feedback.

Our reasoning is based on the following consideration: in order to achieve a pure state at the end of the protocol, either there is a single possible measurement outcome $P(M)=\delta_{M, M'}$ (and thus $H=0$) or the feedback process has to distill the mixed state $\rho_1^M = \sum_M \ket{\psi(M)}\bra{\psi(M)}$ (with $\ket{\psi(M)}_S = \bra{M}_A U_1 \ket{0}_{SA}$) ($H\neq0$) into a pure state through conditional feedback.
It is much easier to fulfill the first condition than the second one.
As a consequence, if the feedback part of the protocol is updated too slowly, it may appear convenient for the optimizer to simply tune $U_1(\theta_1)$ to prepare the best possible state where the ancillas are in a product state and the action of the measurement is irrelevant. 
Otherwise stated, either all the information extracted from the system using the measurement needs to be effectively used to reduce the mixedness of the final state, or else the optimization algorithm will decide to reduce how much the system and ancillas are entangled. 
This would also favor $H$ to settle mostly on integer values of $H$, as it is easier to fully use an ancilla or not use it at all.
To address this issue, the rate of learning of the parameters of the feedback operation can be increased relative to that of the parameters of the initial unitary.
This adjustment can be implemented in several ways: here, we choose to update the parameters of the feedback operation more frequently, as described in Alg.~\ref{alg:update_freq}.
\begin{algorithm}[H]
\caption{Optimization algorithm that updates the feedback parameters $W$ more frequently than the initial unitary parameters $\theta_1$. This approach aims to help the feedback mechanism to adapt more rapidly, preventing the optimization process from ignoring the measurement and feedback steps that lead to local minima.}
\label{alg:update_freq}
\begin{algorithmic}
\Function {update parameters}{$\theta_1$,$W$, loss}
    \State  $\theta_1^\text{new} \gets \text{ADAM\_update}(\theta_1, \nabla_{\theta_1}\text{loss}(\theta_1,W))$
    \State  $W^\text{new} \gets W$
    \For{1 to update\_freq}   
        \State  $W^\text{new} \gets \text{ADAM\_update}(W^\text{new}, \nabla_{W}\text{loss}(\theta_1,W^\text{new}))$
    \EndFor

    \Return $\theta_1^\text{new}, W^\text{new}$
\EndFunction
\end{algorithmic}
\end{algorithm}

\begin{figure}[t]
    \centering
    \subfloat[]{\label{fig:mvqe:freq_ancilla_regu=0:infid}\includegraphics[width=0.47\textwidth]{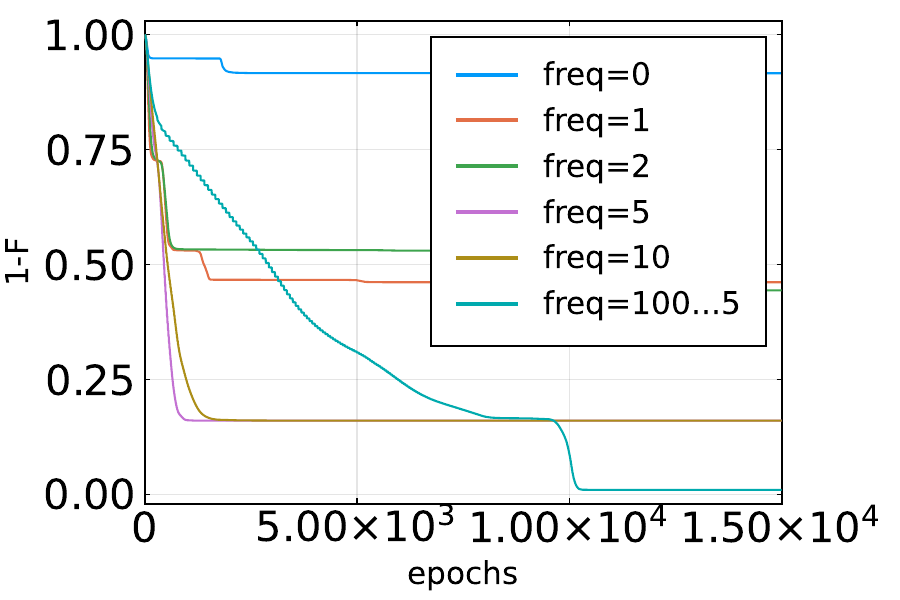}}
    \\ \hspace{0.025\textwidth}
    \subfloat[]{\label{fig:mvqe:freq_ancilla_regu=0:entropy}\includegraphics[width=0.43\textwidth]{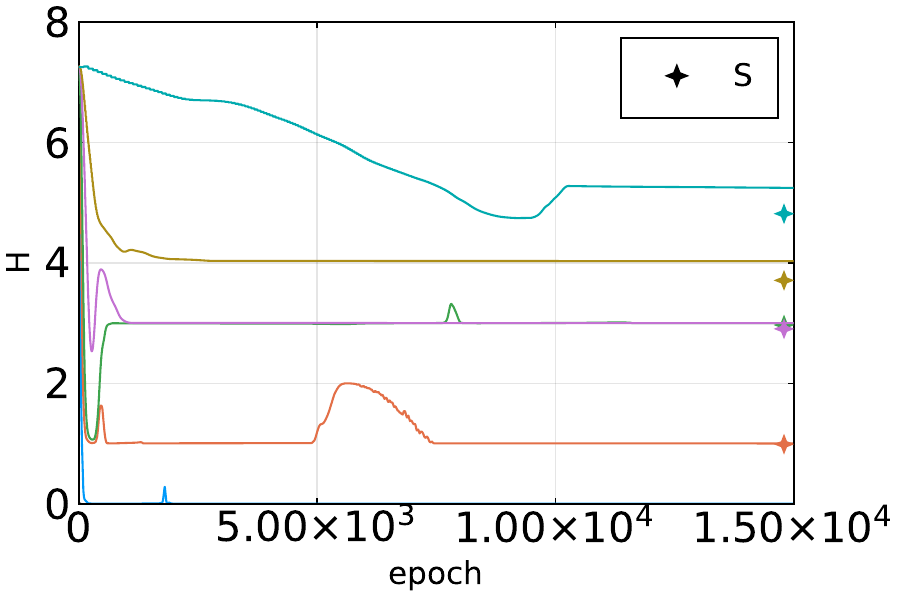}}
    
    \caption{Influence of feedback update frequency on local minima encountered during the optimization of the feedback protocol to prepare any of the four AKLT states with a system size of 16 and 8 ancillas, configured in a repeating ASSSSA pattern where A stands for ancilla and S for the system. Every epoch, the gradient is calculated using all possible measurement outcomes. The parameters of the feedback operation are updated more frequently using gradients. The label "freq=100..5" indicates that the update frequency decreases linearly from 100 to 5 over $10^4$ epochs. (a) Infidelity to the AKLT states: higher update frequencies prevent the protocol from getting trapped in local minima. (b) Shannon entropy $H$ of the measurement probability distribution $P(M)$ and entanglement entropy between the ancilla and the system before measurement at the end of the optimization process (indicated by the star marker): higher update frequencies result in higher Shannon and entanglement entropies. Note that the entanglement entropy between the ancilla and the system was computed at the end of the optimization only once as it is expensive to calculate.
    }
    \label{fig:mvqe:freq_ancilla_regu=0}
\end{figure}%

To test the hypothesis, in Fig.~\ref{fig:mvqe:freq_ancilla_regu=0} the feedback protocol is optimized for the AKLT state with 16 physical qubits and 8 ancilla qubits for different update frequencies of the feedback step. The system size 16 is chosen as it is large enough that every optimization run gets trapped in local minima, making this system size a compelling challenge for analysis.

If the update frequency of the feedback parameters is set to 0 (see blue line in Fig.~\ref{fig:mvqe:freq_ancilla_regu=0}), they are not updated at all. Consequently, the pre-measurement unitaries are forced to prepare an ancilla state that, when measured, does not affect the final state, resulting in $H=0$.

With the standard approach, updated frequency equal to one, the optimization gets trapped in a local minimum with $H=1$, corresponding to one bit of information contained in the ancilla distribution.
The entanglement entropy $S$ between system and ancilla at the end of the optimization is also one, confirming that one bit of information is extracted from the system with the measurement.

Higher update frequencies for the feedback parameters lead to improved infidelities and both higher Shannon entropy $H$ and entanglement entropy $S$, which lie close to each other, showing that our initial conjecture seems to be correct. The low entropy is caused by the pre-measurement unitary trying to bypass the measurement step. And interestingly the Shannon entropy mostly settles to integer values, showing that as expected these are atractors of the optimization dynamics.

The most effective protocol with this strategy is achieved by initially setting the update frequency to 100 and then linearly decreasing it to 5 over the first $10^4$ epochs achieving the highest value of $H$ and showing that high update frequencies are mostly important at the beginning of the optimization process. The optimal update frequency should be chosen such that the feedback operation parameters are close to $\min_W \text{loss}(\theta_1, W)$.

\subsection{Ancilla Regularization}
\label{sec:mvqe:ancilla_regu}

Another approach to prevent the measurement distribution from becoming trapped in local minima, characterized by $H = 0$, involves encouraging broader exploration of the solution space. This is achieved by regularizing the distribution $P(M)$, by adding a term to the loss function that encourages a more uniform probability distribution. The regularization term $l_R(M; \theta)$ is defined as follows:
\begin{align}
d(M; \theta) &= 1 + \frac{\log_2(P(M; \theta))}{N_a}\\
l_R(M; \theta) &= 
\begin{cases} 
0 & \text{if } -c < d < c \\
(d(M; \theta)+c)^2 & \text{if } d < -c \\
(d(M; \theta)-c)^2 & \text{if } c < d 
\end{cases}\\
l_R(\theta) &= \frac{1}{2^{N_a}}\sum_M l_R(M; \theta)
\end{align}
where the distance $d(M)$ tells us how far the measurement sample $M$ is from having the probability $2^{-N_a}$ where $N_a$ is the number of ancillas. The window width $c$  was chosen so that if $l_R = 0$ then $\frac{\max_M P(M)}{\min_M P(M)} < r$ with $c = \frac{\log_2(r)}{2N_a}$.  In our experiments, the ratio $r = 2$ is chosen, ensuring that when $l_R(\theta) = 0$, the largest and smallest measurement probabilities differ by a factor of at most two. This promotes a near-uniform distribution of measurement results, mitigating the risk of the protocol becoming trapped in low-$H$ regions that cannot effectively utilize feedback.

\begin{figure}[t]
    \centering
    \subfloat[]{\label{fig:mvqe:freq_ancilla_regu=1:infid}\includegraphics[width=0.47\textwidth]{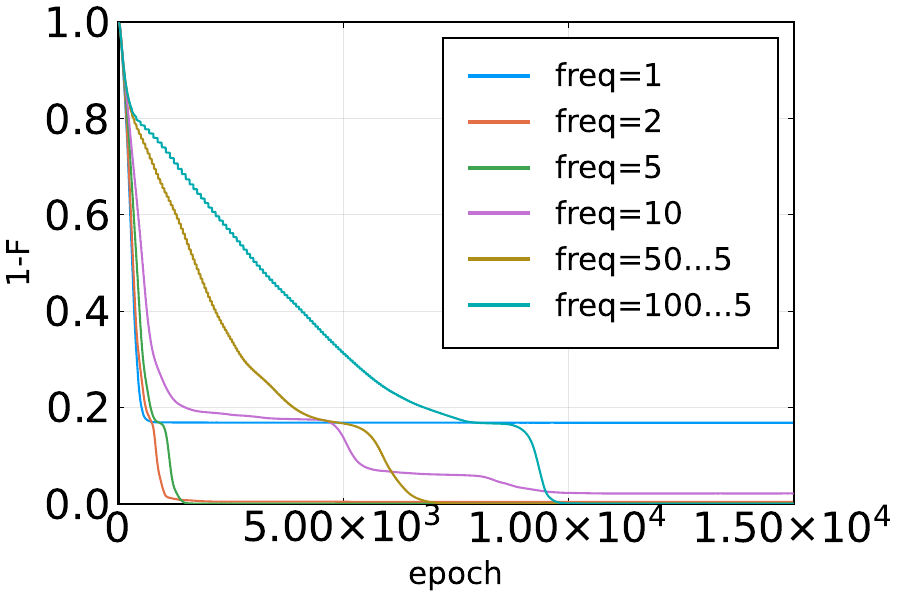}}
    \\ \hspace{-0.02\textwidth}
    \subfloat[]{\label{fig:mvqe:freq_ancilla_regu=1:entropy}\includegraphics[width=0.49\textwidth]{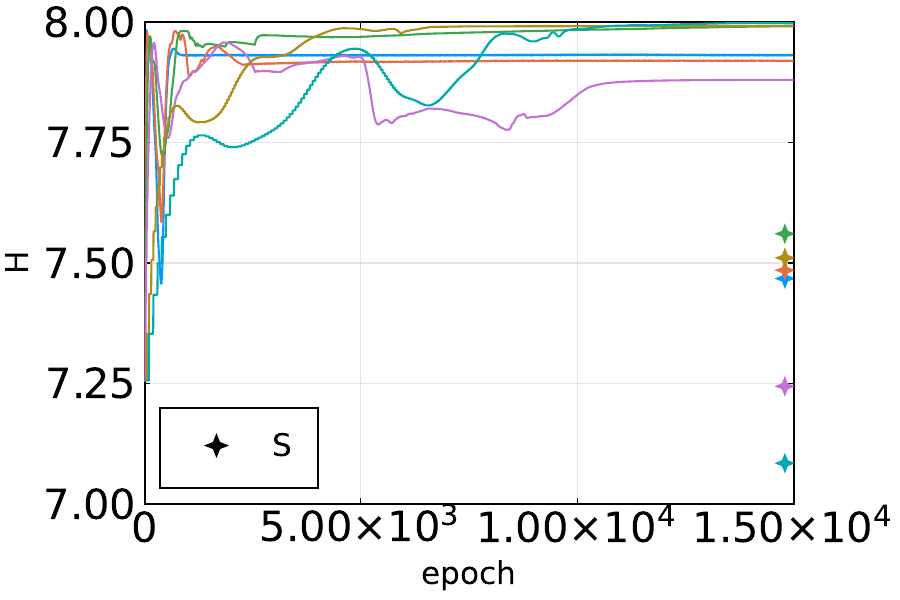}}
    
    \caption{Same optimization like in Fig.~\ref{fig:mvqe:freq_ancilla_regu=0} but with the ancilla regularization turned on.}
    \label{fig:mvqe:freq_ancilla_regu=1}
\end{figure}

Fig.~\ref{fig:mvqe:freq_ancilla_regu=1} shows the effect of adding the regularization term to the loss function during optimization. 
With the regularization enabled, the Shannon entropy directly takes its maximal value of $H=N_a$.
With a feedback update frequency of 1 the infidelity reaches a local minimum with lower infidelity than without regularization.

When the update frequency is doubled to 2 or higher, entrapment in local minima occurs less frequently, showing the effectiveness of the regularization procedure.
However, even though less frequently and at lower infidelities local minima can still appear. The optimization with an update frequency of 10 in Fig.~\ref{fig:mvqe:freq_ancilla_regu=1:infid} is a good example. The reason for the improvement can be correlated to the high Shannon and entanglement entropies reached due to the regularization.

Note that choosing a large update frequency does not provide significant additional benefits in avoiding local minima, but it does cause the optimization to take much longer to converge.

\section{Comparing learned and reference protocol}
\label{sec:mvqe:comparison}

\begin{figure*}[ht]
    \centering
    \subfloat[]{\label{fig:mvqe:girvinMI:U}\includegraphics[height=0.3\textwidth]{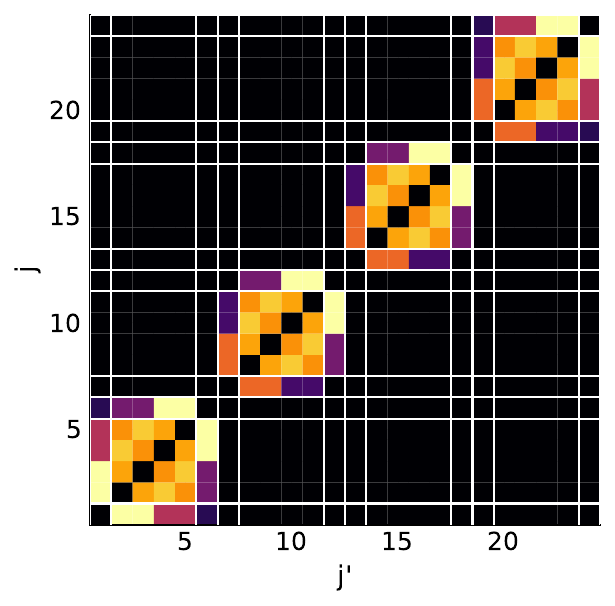}}
    \subfloat[]{\label{fig:mvqe:girvinMI:UM}\includegraphics[height=0.3\textwidth]{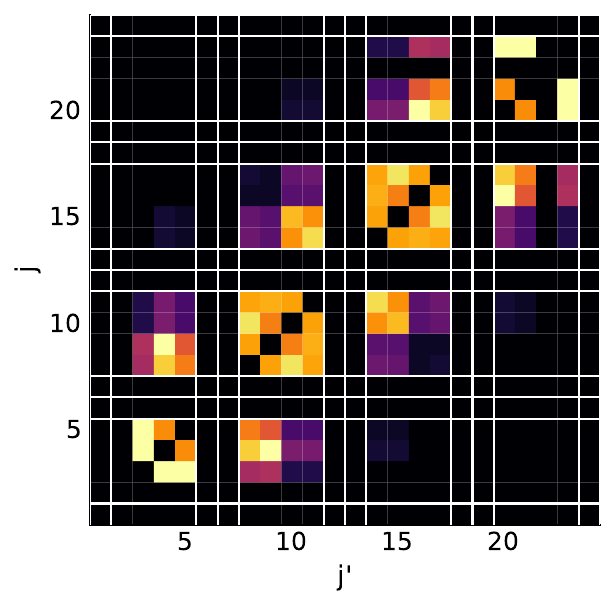}}
    \subfloat[]{\label{fig:mvqe:girvinMI:UMC}\includegraphics[height=0.3\textwidth]{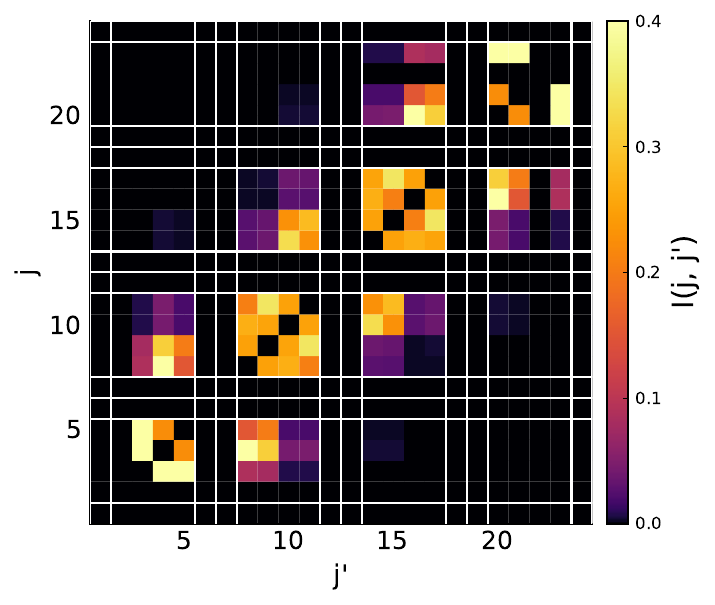}}
    \\
    \subfloat[]{\label{fig:mvqe:learnedMI:U}\includegraphics[height=0.3\textwidth]{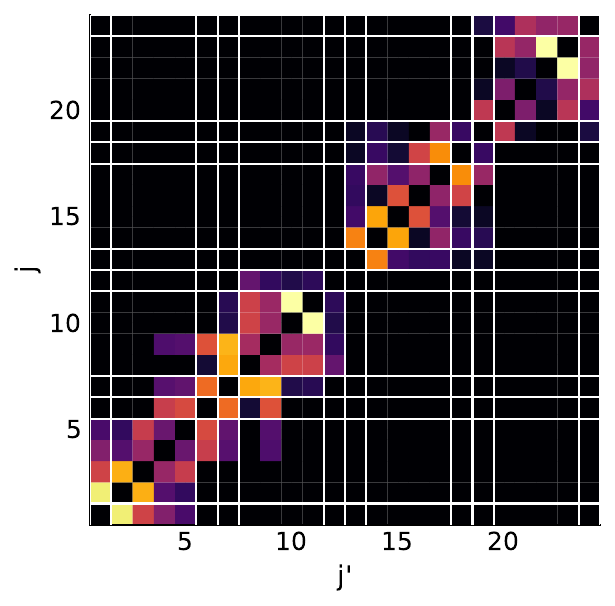}}
    \subfloat[]{\label{fig:mvqe:learnedMI:UM}\includegraphics[height=0.3\textwidth]{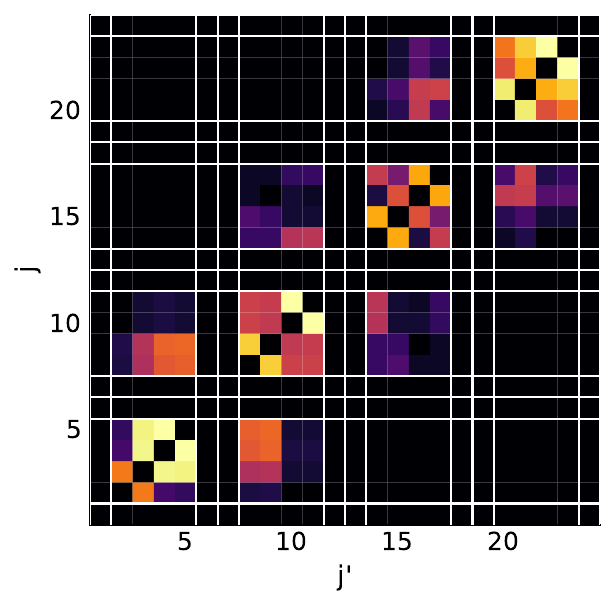}}
    \subfloat[]{\label{fig:mvqe:learnedMI:UMC}\includegraphics[height=0.3\textwidth]{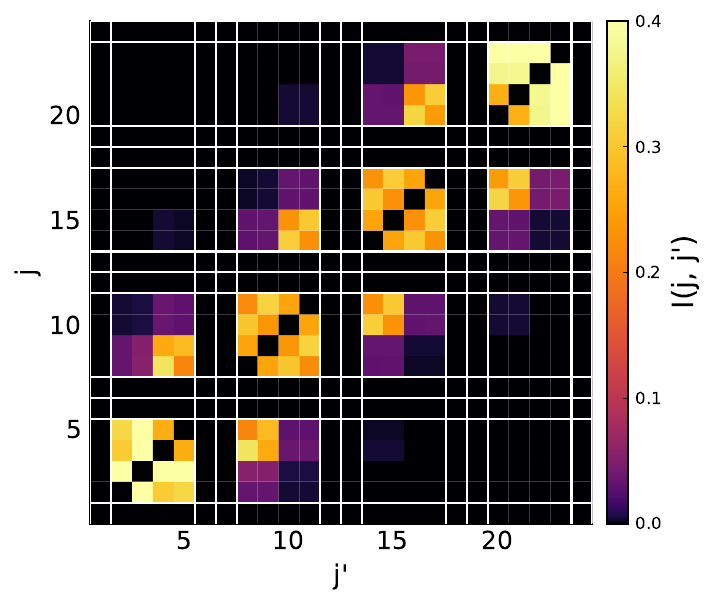}}\\
    \subfloat[]{\label{fig:mvqe:learnedMI:distance}\includegraphics[width=0.41\textwidth]{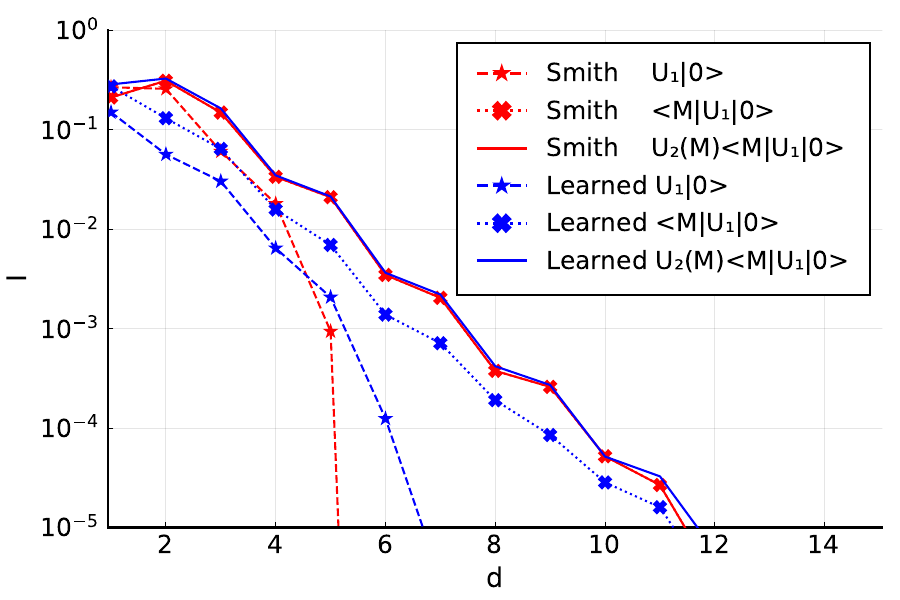}}
    
    \caption{Mutal Information $I(j, j')$ between different sites of the quantum state after the three operations in the protocol for the analytically derived protocol (a,b,c) and the learned protocol (d,e,f), where the white lines signify that the corresponding qubit is an ancilla. First, both ancillas and system are initialized in the $\ket{0}$ state and the first unitary is applied $U\ket{0}_A\ket{0}_S$ (a, d). Then a measurement is performed $\bra{M}_A U\ket{0}_A\ket{0}_S$ (b, e). Finally, the resulting state is corrected conditionally on the measurement outcome $U_S(M)\bra{M}_A U_{SA}\ket{0}_S\ket{0}_A$. (g) Is the averaged mutual information $\bar{I}(d)=\frac{1}{N(d)}\sum_{j'}I(j,j'\pm d)$ at distance $d$ for both the analytically derived and learned protocol.}
    \label{fig:mvqe:girvinMI}
\end{figure*}%

In this section, the similarities and differences between the learned and the analytically derived protocol by Smith~\etal~\cite{smith2023deterministic} are examined. 
The primary tool for characterizing the feedback protocol is the analysis of intermediate states using quantum mutual information between two qubits:
\begin{align}
I(j,j')=&S(\rho_i^R) + S(\rho_{j'}^R) - S(\rho_{j, j'}^R)\\
\rho_j^R =& \mathrm{Tr}_{\bar{j}} (\ketbra{\psi}{\psi}),
\end{align}
where $\mathrm{Tr}_{\bar{j}}$ traces out the entire system other than the j-th qubit. This quantity measures the extent to which two qubits are entangled when the rest of the system is traced out.
Furthermore, we examine whether the learned feedback on a certain qubit pair only depends on the measurement outcomes of the qubits on its left/right correctability.

The analytically derived protocol begins by preparing small AKLT states whose boundaries are entangled with adjacent ancillas. 
This is shown in Fig.~\ref{fig:mvqe:girvinMI:U}, where the mutual information of that state reveals 6x6 blocks of strong entanglement. 
The ancillas are then measured in the Bell basis (see Fig.~\ref{fig:mvqe:girvinMI:UM}), where the boundary conditions of the small AKLT states are merged. 
This process introduces defects into the AKLT state that depend on the measurement results.
These defects, which can not be seen in the mutual information, can then be corrected during the feedback step (see Fig.~\ref{fig:mvqe:girvinMI:UMC}), allowing the system to randomly reach any of the four AKLT states.

The learned approach, on the other hand, is not limited to the block structure of the analytically derived protocol. While it learns a similar structure, it exhibits greater flexibility by generating entanglement beyond the 6x6 sub-lattice. This additional flexibility can become particularly valuable when preparing other quantum states, as the learned protocol can exploit this ability to improve its performance.

Notably, the learned protocol needs to generate less mutual information prior to measurement (see Fig.~\ref{fig:mvqe:learnedMI:U}) compared to the analytically derived one.
The analytically derived protocols feedback gates are restricted to the creation or removal of excitations in the AKLT state in the feedback step.
A comparable mutual information pattern emerges when the learned protocol is constrained to the same correction gates.
Interestingly, this shows not only that there are many (almost equivalent) variants of the protocol, but also that most of them require less mutual information before measurement than the analytically derived protocol.

In Fig.~\ref{fig:mvqe:learnedMI:distance}, both the learned and analytically derived protocol initially exhibit a maximum mutual information length, $\bar{I}(d)$, of 6 before measurement, reflecting the presence of well-structured, localized patches of mutual information generated by short-range unitaries. After measurement, the mutual information decays exponentially, similar to the AKLT state. This behavior demonstrates how both protocols effectively fuse these localized patches into a unified state, transforming short-range correlations into exponentially decaying ones.

Finally note that the smaller mutual information is consistent across all learned protocols, making them easier to implement using hardware-efficient approaches at a smaller circuit depth of 7. If, instead of learning the protocol from scratch, one attempts to replicate the analytically derived protocol by maximizing $f(\theta)=\bra{0}U_1^{\text{smith}\dag}U_1^\text{learned}(\theta)\ket{0}$, a circuit depth of 8 is required. Note that the minimum theoretical circuit depth to entangle the system S and ancilla A arranged in the pattern ASSSSA is 6.

Another aspect of interest is whether the feedback is left- or right-correctable. This property implies that the feedback parameters $\theta_2^{i,j}=f(M_1, \ldots, M_{j-1})$ depend only on measurements from one side. For example, this is the case for the analytically derived protocol of Smith~\etal ~\cite{smith2023deterministic}. To test if the learned protocol is left correctable, we examine how the pre-measurement unitary $U_1(\theta_1)$, which is initially learned without any constraints on left correctability, can be corrected with left correctable feedback. We do this by constraining a new feedback unitary $U_2(\theta'_2)$ to be learned under the requirement of being left correctable (see~App.~\ref{app:mvqe:leftcorrectable}). The results indicate that $U_1$ is indeed left correctable, even though no such constraint is explicitly applied during its training. Furthermore, this is the case for all protocols found in this work.

\section{Periodic Ansatz \& RNN}
\label{sec:mvqe:PeriodicAnsatzRNN}
\begin{figure*}[t]
    \subfloat[]{\label{fig:mvqe:rnn:Ns=16}\includegraphics[width=0.33\textwidth]{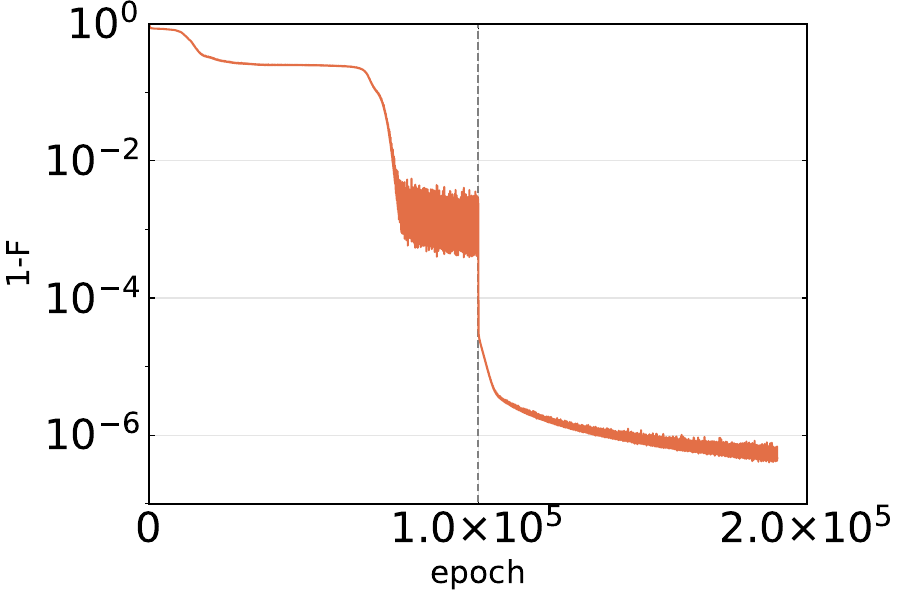}}
    \subfloat[]{\label{fig:mvqe:rnn:statsize}\includegraphics[width=0.33\textwidth]{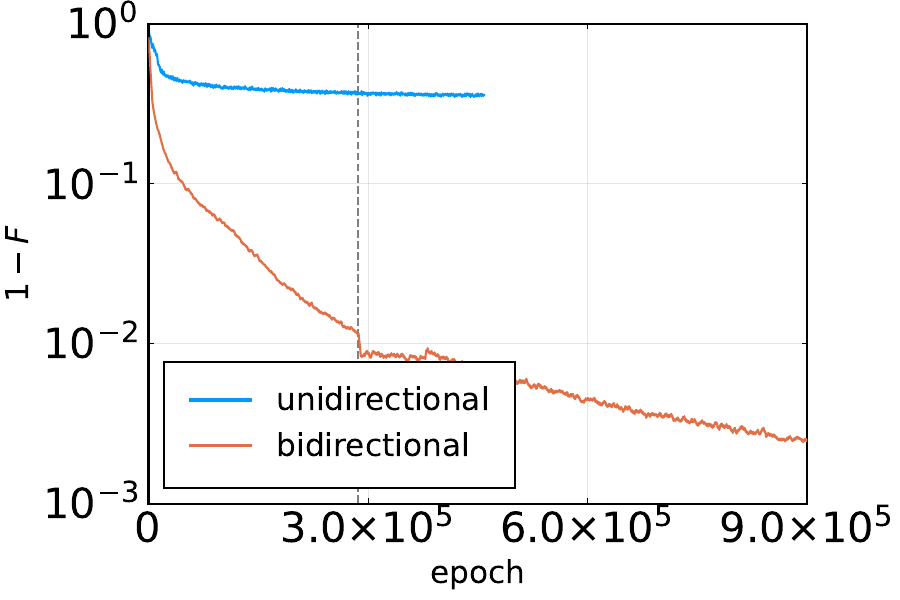}}
    \subfloat[]{\label{fig:mvqe:rnn:performance}\includegraphics[width=0.33\textwidth]{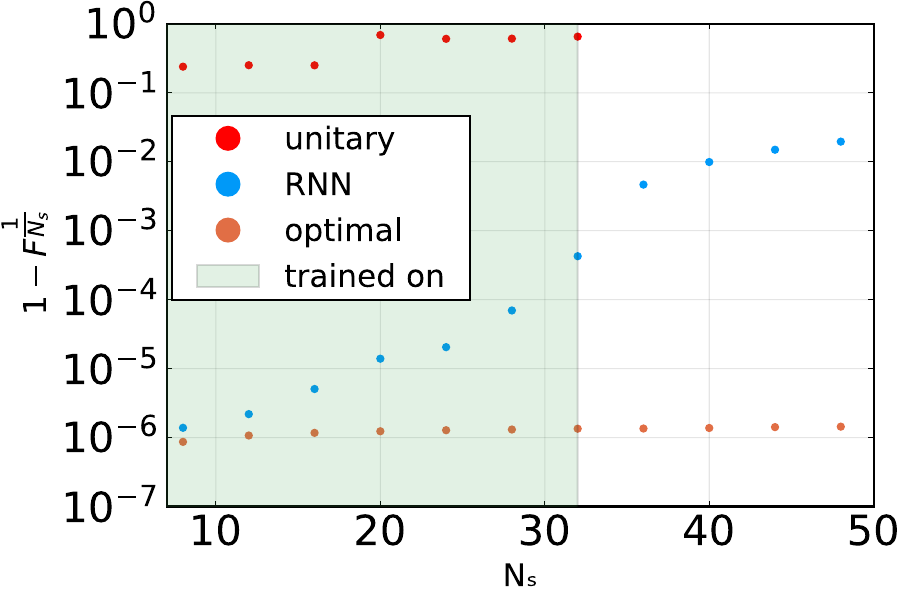}} 
    
    \caption{Figure (a) presents the optimization of the protocol for a system size of $N_s = 16$, utilizing a translationally invariant ansatz for the pre-measurement unitary $U_1$, with an RNN serving as the feedback function. The ADAM optimizer’s learning rate was reduced from $10^{-3}$ to $10^{-5}$ after $10^5$ epochs. Figure (b) shows the RNN performance following re-initialization, with the angles of the pre-measurement unitary frozen. The RNN was trained concurrently on system sizes $N_s = [8, 12, 16, 20, 24, 28, 32]$, and the curve was smoothed with a running mean as the raw data was noisy. The blue curve shows the training for a unidirectional RNN and the orange curve for a bidirectional RNN. The gradient is calculated every epoch for 144 measurement outcomes. In Figure (c), instead of the infidelity the infidelity per site $1-F^{1/N_s}$ is plotted to be able to compare system sizes meaningfully. The red points indicate the infidelity reached when only optimizing unitaries, removing the ancillas and measurements, the blue points indicate the performance of the RNN feedback on 1,000 measurement outcomes. In contrast, the orange points show further optimization of the feedback angles by minimizing the infidelity until convergence only for the 1,000 measurement outcomes. This comparison reveals that although the RNN does not learn the optimal feedback strategy, it performs well for small system sizes, and the learned strategy successfully extrapolates to larger sizes, even though better performance was hoped for.}
    \label{fig:mvqe:rnn}
\end{figure*}%

The ansatz presented in this section seeks to autonomously learn a translationally invariant strategy with the help of Recurrent Neural Networks (RNN) to prepare the four-fold AKLT state manifold. The advantage of this approach lies in its ability to be trained on a set of fixed system sizes and subsequently extrapolated to larger ones. 
This ansatz builds upon the methods introduced in previous sections but incorporates two key modifications.
First, the ansatz is modified so that the angles of the pre-measurement unitary are set to repeat with a periodicity of six, i.e., $\theta_{i,j} = \theta_{[i,j \text{ mod } 6]}$. This is a natural choice as the qubits repeat in the ancilla qubit A and system qubit S pattern ASSSSA.
Second, the feedback mechanism is designed to be system-size independent by replacing the tabular feedback approach with an RNN $f$. 
The RNN architecture comprises five layers, alternating between Gated Recurrent Units (GRUs) to propagate measurement information spatially and Swish-Gated Linear Units (SwiGLUs) to process information locally. For these experiments, we selected a hidden dimension of $d_h = 60$. Additional architectural details can be found in App.~\ref{appendix:mvqe:rnn}.

The optimization procedure for $N_s=16$, combining the translationally invariant pre-measurement unitary $U_1$ with bidirectional RNN-based feedback, is illustrated in Fig.~\ref{fig:mvqe:rnn:Ns=16}. 
Using the translationally invariant ansatz introduces local minima that trap the optimization process. 
To mitigate this, the optimization was repeated three times until a successful run avoided these local minima. Once the process escapes the local minima, further optimization requires only adjusting the learning rate when necessary. The procedure was halted upon reaching a sufficiently low infidelity of $10^{-6}$.

However, the protocol at this stage is limited to the system size $N_s=16$ since the RNN has not yet generalized to other sizes. To address this, we freeze the pre-measurement unitary angles and reinitialize the RNN. The RNN was then trained simultaneously on a range of system sizes, $N_s = [8, 12, 16, 20, 24, 28, 32]$, using the following loss function:
\begin{align}
\ket{\psi_{N_s}(M)} &= \bra{M}_AU_1(\theta_1)\ket{0}_S^{\otimes N_s}\ket{0}_A^{\otimes N_a}\\
P_{N_s}(M) &= \brakets{\psi_{N_s}(M)}{\psi_{N_s}(M)} \\
F(M, N_s; W) &= \frac{|\bra{\psi_{\text{target}}}U_2(\theta_2=f(M;W))\ket{\psi_{N_s}(M)}|^2}{P_{N_s}(M)}\\\
\text{loss}(\theta) &= 1 - \frac{1}{\mathcal{N}(N_s)} \sum_{N_s} \langle F(M, N_s; W)^{1/N_s} \rangle_{M \in P_{N_s}(M)}.
\end{align}
This loss function averages over sampled measurement results at different system sizes, and uses the fidelity per site in order to avoid system size artifacts.

Initially, a unidirectional RNN was employed, as analysis using the algorithm in App.~\ref{app:mvqe:leftcorrectable} indicated that the protocol was left-correctable. However, as shown in Fig.~\ref{fig:mvqe:rnn:statsize}, the unidirectional RNN yielded poor performance. Surprisingly, no clear explanation has been identified for this underperformance. 
Only after switching to a bidirectional RNN did the model achieve satisfactory results, with average infidelity across all trained system sizes reducing to $3 \cdot 10^{-3}$ after $9 \cdot 10^{5}$ epochs. 
During training, as a test, the learning rate was adjusted to a cosine schedule~\cite{loshchilov2016sgdr} at $2.9 \cdot 10^5$ epochs, which further improved performance and was kept as a result. 
The optimization was ultimately halted at epoch $9\cdot 10^{5}$ even though the loss was still decreasing as the training time became prohibitively long. At the last epoch, the infidelity at each system size was assessed individually and plotted in Fig.~\ref{fig:mvqe:rnn:performance}.

In this analysis, we use the infidelity per site, defined as $1 - F^{1/N_s}$, instead of the total infidelity as the figure of merit. This choice reflects the exponential scaling of fidelity with system size and allows for a more meaningful comparison across different $N_s$. For example, even when preparing a product state using single-site unitaries a modest constant per-site error will lead to a fidelity that decays exponentially with $N_s$, $F=(1-\epsilon)^{N_s}$. Using fidelity when comparing different system sizes would mask the actual per-site performance of the protocol. Note here that for small $\epsilon$ the infidelity grows linearly with system size $1-F=1-(1-\epsilon)^{N_s}\approx \epsilon N_s$. The resulting protocol whose performance can be seen in Fig.~\ref{fig:mvqe:rnn:performance}, prepares states with infidelities that grow exponentially with system size and the performance deteriorates faster than expected for larger system sizes. 

To better understand what is causing the poor performance the correction angles were further optimized for 1,000 values of $M$ to find the best possible feedback for them.

It is observed that the infidelity per site for the optimal correction hardly scales with system size. This indicates that the RNN could still be improved, but nonetheless, it gives good performance for system sizes that are small enough. In Fig.~\ref{fig:mvqe:rnn:statsize} one can see that the average infidelity is still decreasing, but due to the expense of the gradient calculations and the many steps required, the RNN was not optimized until convergence.

The slow convergence could be due to vanishing gradients caused by the RNN architecture. It is possible that other architectures like Transformers~\cite{vaswani2017attention} or Mamba~\cite{gu2023mamba}, which have better gradient flow could learn faster. This is left for future investigations.

\section{Other states/ a specific AKLT state}
\label{sec:mvqe:SpecificAKLT}

\begin{figure*}[t]
    \centering
    \subfloat[]{\label{fig:mvqe:oneakltMI:training}\includegraphics[width=0.4\textwidth]{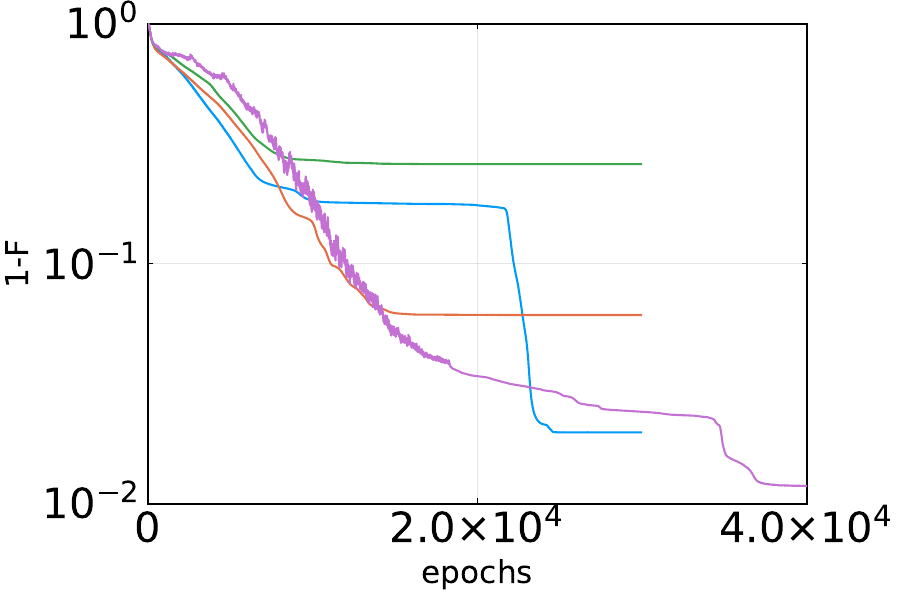}} \\
    \subfloat[]{\label{fig:mvqe:oneaklt:U}\includegraphics[height=0.3\textwidth]{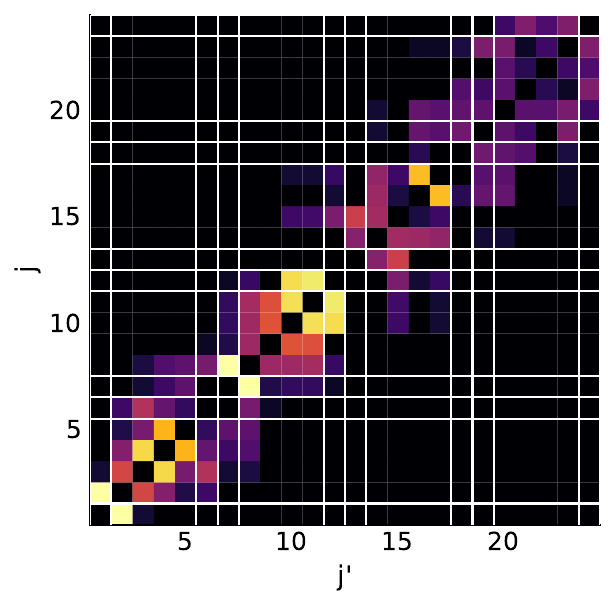}}
    \subfloat[]{\label{fig:mvqe:oneaklt:UM}\includegraphics[height=0.3\textwidth]{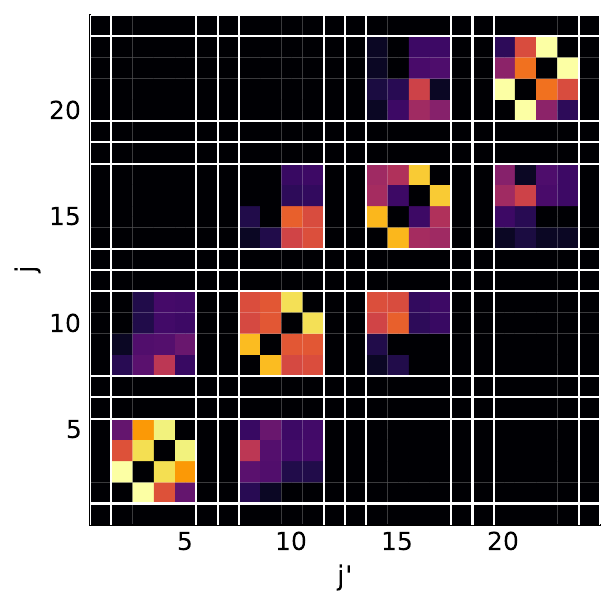}}
    \subfloat[]{\label{fig:mvqe:oneaklt:UMC}\includegraphics[height=0.3\textwidth]{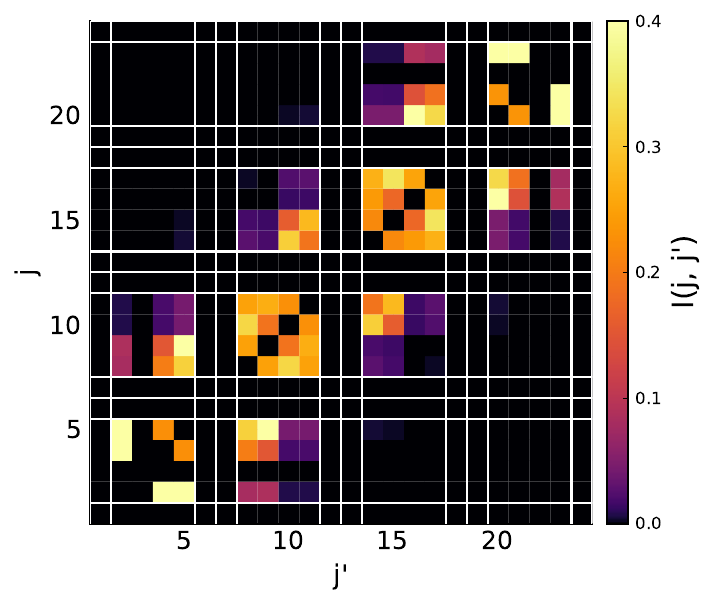}} 
    
    \caption{(a) Infidelity evolution during protocol optimization for preparing the AKLT state with both edge modes in the spin-up state. Blue, orange, and green lines represent optimizations with different initial seeds. The violet line shows a particularly successful "lucky" run that achieved significantly lower infidelity, likely due to finding a more optimal path in the parameter space. This demonstrates the potential for high-fidelity state preparation when favorable parameters are found.
    (b-d) Mutual Information $I(j, j')$ between different sites of the quantum state after each operation in the learned protocol. White lines indicate ancilla qubits.
    (b) After applying the first unitary: $U_1\ket{0}_A\ket{0}_S$, where both ancillas and system are initialized in the $\ket{0}$ state.
    (c) After measurement: $\bra{M}_A U_1\ket{0}_A\ket{0}_S$.
    (d) After the feedback step: $U_2(\theta_2=f(M;W))\bra{M}_A U_1\ket{0}_S\ket{0}_A$, resulting in the prepared AKLT state with spin-up edge modes.}
    \label{fig:mvqe:oneaklt}
\end{figure*}%

In this section, we shift the focus from the random preparation of AKLT states to the consistent production of a specific AKLT state. 
Without loss of generality, our target is the AKLT state with both edge modes in the spin-up configuration. 
This is a challenging task as there is no known deterministic short-circuit protocol for this purpose.

The study of this particular state preparation serves two important purposes. 
Firstly, it demonstrates the feasibility of deterministic state preparation using feedback mechanisms. Secondly, it highlights the potential for discovering novel protocols through learning-based approaches. 
Currently, the Smith~\etal protocol remains the most effective method for this operation, achieving the desired spin-up edge mode state with a 50\% probability, through repeated AKLT state preparations. This is the case as initially each of the 4 AKLT edge states can be prepared with 25\% probability and the edges can easily flipped simultaneously like $\ket{\uparrow\uparrow} \leftrightarrow \ket{\downarrow\downarrow}, \ket{\downarrow\uparrow} \leftrightarrow \ket{\uparrow\downarrow}$. 

We optimize the protocol for the preparation of a single AKLT state using the methodology described in Sec.~\ref{sec:mvqe:ancilla_regu}. Fig.~\ref{fig:mvqe:oneakltMI:training} shows that the optimization process often gets trapped in local minima, as evidenced by the different results of different random seeds. 
To mitigate this issue, alternative strategies were investigated; however, these approaches did not yield significant improvements in the final fidelities. Interestingly, one optimization run achieves low infidelity due to a favorable random seed (see the violet line in Fig.~\ref{fig:mvqe:oneakltMI:training}). This "lucky" run was obtained by adding gradually decreasing random noise to the parameters ($\theta_1,W$) during the optimization process. This demonstrates the potential for high-fidelity state preparation when advantageous parameter configurations are found nad the importance of further improving optimization techniques. Consequently, this "lucky" protocol warrants further analysis.

Fig.~\ref{fig:mvqe:oneaklt:U} shows the mutual information of the intermediate state before the ancilla measurement in the "lucky" protocol. 
The mutual information shows a distinctive behavior compared to the one observed in previous sections. 
In particular, the right side of the chain shows weak mutual information before the measurement.
Fig.~\ref{fig:mvqe:oneaklt:UM} shows that entanglement does not fully propagate through the chain until after the measurement.
This observation suggests a greater reliance on measurements for entanglement propagation in this protocol than for the simpler optimization objective of the previous sections. 
The difference between the pre- and post-measurement mutual information distributions highlights the critical role of quantum measurement in this more complex protocol.

The protocol was found to be left/right correctable which is surprising. The random preparation of any of the 4 AKLT states was expected to be left correctable as the analytically derived protocol is left correctable. But for this protocol, we expected that the feedback gates would need to be dependent on both the left and right sides.

\section{Conclusion}
A self-learning protocol for quantum state preparation is presented that integrates measurement and feedback in variational quantum circuits. 
By incorporating projective measurements and conditional feedback, the protocol learns efficient state preparation strategies beyond unitary-only methods, especially where measurement-based shortcuts reduce circuit depth.

Using the 1D spin-1 AKLT states as a benchmark, the protocol successfully learned a feedback mechanism to prepare these states with high fidelity. 
Notably, the learned approach required less mutual information prior to measurement than the analytically derived protocol of Smith \textit{et al.}, indicating a potential for shallower circuits.

To address local minima during optimization, two strategies were implemented: adjusting parameter update frequencies to balance learning rates between the initial unitary and feedback operations, and introducing ancilla regularisation to promote uniform measurement results. 
These strategies effectively mitigated local minima and improved the performance of the feedback mechanism.

The protocol was extended to larger systems using a translationally invariant approach and a recurrent neural network (RNN) for feedback. 
While the RNN did not fully capture optimal corrections for large systems, it performed well for smaller sizes and showed some potential for generalization, even though better extrapolation performance was expected.

The preparation of a specific AKLT state with both edge modes in the spin-up configuration was also explored - a task for which no known deterministic, low-depth protocol exists. 
The results demonstrated the possibility of learning such a protocol, highlighting the potential of the approach to discover new state preparation methods.

Moreover, a significant potential application of our work lies in addressing open questions in the classification of phases of matter via finite-depth unitaries and feedback. Previous works, such as \cite{tantivasadakarn2023hierarchy,piroli2024approximating}, have introduced the concept of classifying quantum phases using measurement and feedback protocols. However, certain questions remain unresolved—for instance, whether all two-dimensional topological phases become trivial when feedback is added. It is suspected that some, like the Fibonacci anyon phase, remain non-trivial even with feedback. Our algorithm provides a framework that could be used to explore these questions numerically.

A practical consideration for implementing our protocol regards the overhead introduced by measurement and classical feedback. While our simulations assume ideal conditions with instantaneous feedback, current hardware presents significant latency challenges. Active research is addressing this challenge, particularly in the context of quantum error correction \cite{acharya2024quantum,caune2024demonstrating,sivak2023real}. For instance, in a recent experiment on real-time quantum error correction \cite{acharya2024quantum}, Google reports a QEC cycle time of 1.1~$\mu$s, with a real-time decoder achieving an average feedback latency of 63~$\mu$s—corresponding to a backlog of approximately 57 cycles. These constraints currently limit the applicability of feedback-based protocols to small system sizes or require buffering and delayed correction strategies. Nevertheless, the rapid pace of experimental progress suggests that these latencies will be reduced in the near future, enabling real-time feedback as envisioned in our approach.

In conclusion, the incorporation of measurement and feedback into variational quantum algorithms offers a promising framework for quantum state preparation. 
This approach addresses optimization challenges but does not completely solve them and extends quantum state preparation tools by exploiting the non-unitary effects of measurements. 
Future work should focus on using these learning techniques to design protocols with multiple rounds of feedback since most analytical protocols rely on only one round of measurements, and we see potential in the multi-round approach, which we imagine will be particularly useful in the presence of errors. 
Another important extension of this work is the implementation of such a protocol in an experiment, once measurement-based feedback becomes available. We envision further fine-tuning the measurement protocol in the experiment using reinforcement learning techniques so that it can learn to mitigate errors present on real hardware.

\textit{Code and simulation data—}  
All simulations were performed using the MPS formalism to represent the wave function; the code and resulting data are available on Zenodo~\cite{zenodo}. The Julia library mVQE.jl~\cite{mVQE}, developed specifically for these simulations, uses the ITensors.jl~\cite{ITensors} library as its backbone.

{\it Note added.—} During the final stages of preparing this manuscript, we became aware of related work by Alam~\etal~\cite{alam2024learning} and Yan~\etal~\cite{ yan2024variational}, which employs a variational ansatz similar to the one presented here. Additional details on the distinctions between the two approaches have been included in the \hyperref[sec:alam_aknowledgement]{Introduction}.

\subsection*{Acknowledgements}
We thank Giovanna Morigi and Lorenzo Piroli for inspiring discussions on the topic. We acknowledge funding by the German Federal Ministry of Education and Research (BMBF) for support under the thematic programme ``Quantum technologies -- from the basics to the market'', project number 13N16202 ``Noise in Quantum Algorithms (NiQ)''. 
This work was also partially funded by the Deutsche Forschungsgemeinschaft (DFG, German Research Foundation) via Project-ID 277101999 -- CRC network TRR 183 (``Entangled states of matter'') and under Germany’s Excellence Strategy – Cluster of Excellence Matter and Light for Quantum Computing (ML4Q) EXC 2004/1 – 390534769.
The authors gratefully acknowledge the Gauss Centre for Supercomputing e.V. (www.gauss-centre.eu) for funding this project by providing computing time through the John von Neumann Institute for Computing (NIC) on the GCS Supercomputer JUWELS at the J\"ulich Supercomputing Centre (JSC) (Grant NeTeNeSyQuMa) and the FZ J\"ulich for computing time on JURECA (institute project PGI-8)~\cite{JURECA2021}. 
%

\bibliographystyle{quantum}
\bibliography{literature.bib}



\newpage

\appendix

\onecolumngrid
\section{The Greenberger–Horne–Zeilinger State}
\label{appendix:mvqe:ghz}
\begin{figure*}[hbt]
    \centering
    \label{fig:mvqe:ghz:stat}\includegraphics[width=0.45\textwidth]{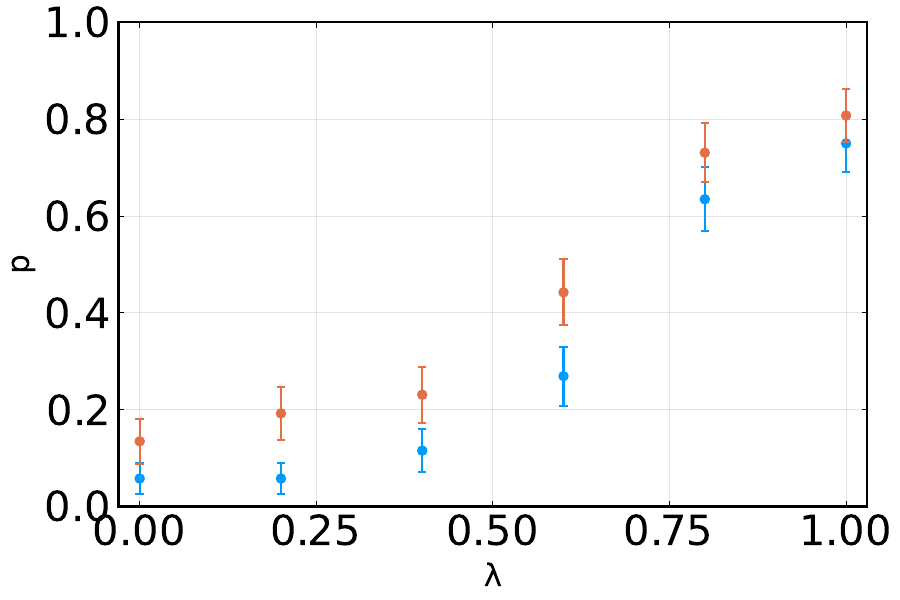} 
    
    \caption{Success probability $p$ as a function of the regularization parameter $\lambda$ for two different optimization methods for the preparation of the 6 qubit GHZ state. The blue data points correspond to the standard ADAM optimizer, while the orange data points represent the optimizer with ancilla regularization and an update frequency of 5. The error bars indicate the standard deviation over 50 independent optimization runs. As $\lambda$ increases, the success probability improves for both methods, with the ancilla-regularized approach consistently achieving higher success rates across all values of $\lambda$. 
    }
    \label{fig:mvqe:ghz}
\end{figure*}%

To validate our claims regarding local minima, we analyzed the simpler GHZ state, for which an efficient feedback protocol is known \cite{watts2019exponential}. Due to its simplicity, the optimization process was significantly faster compared to the AKLT state, allowing us to perform statistical analysis on the optimization methods presented in this work. The infidelity used to optimize the GHZ state preparation protocol is defined as:
\begin{align}
l(\theta;M)&=1-F(\theta; M) \\
&=1-\frac{1}{2}\big\vert (\bra{00..}+\bra{11..}) \ket{\psi(\theta; M)} \big\vert ^ 2 \nonumber\\
&= 1-\frac{1}{2}\big\vert \braket{00..|\psi(\theta; M)} \big\vert ^ 2 - \frac{1}{2}\big\vert \braket{11..|\psi(\theta; M)} \big\vert ^ 2 - \operatorname{Re}
 \left( \braket{00..|\psi(\theta; M)} \braket{\psi(\theta; M)|11..} \right) \; . \nonumber
\end{align}
The first two terms in the fidelity introduce two local minima corresponding to the trivial product states $\ket{00..}$ and $\ket{11..}$. The last term reaches its minimum at the GHZ state, $\text{argmin}_\psi\operatorname{Re} \left( \braket{00..|\psi} \braket{\psi|11..} \right)=\ket{\text{GHZ}}$. As discussed in previous sections, measurement-based variational quantum circuits often encounter difficulties due to local minima induced by underutilization of the ancilla qubits. If these two local minima are not suppressed, finding the target state becomes challenging. To address this issue, we introduce a parameter $\lambda$ to attenuate the impact of the local minima and redefine the loss function as follows:

 \begin{align}
 l(\theta; M) = \left(1 - \frac{\lambda}{2}\right) -\frac{1-\lambda}{2}\big\vert \braket{00..|\psi(\theta; M)} \big\vert ^ 2 - \frac{1-\lambda}{2}\big\vert \braket{11..|\psi(\theta; M)} \big\vert ^ 2 - \operatorname{Re}
 \left( \braket{00..|\psi(\theta; M)} \braket{\psi(\theta; M)|11..} \right) \; .
 \end{align}

For $\lambda = 0$, the loss function reduces to the standard infidelity, while for $\lambda = 1$, it exclusively retains the last term. We conducted 50 optimization trials for various values of $\lambda$ using a small system size of $N_s=6$, where the optimizations were completed within a few minutes. Both a standard ADAM optimizer and an enhanced version incorporating ancilla regularization with an update frequency of 5 were employed. As expected, higher values of $\lambda$ resulted in fewer occurrences of local minima, and the proposed local minima avoidance strategies improved convergence by reducing the likelihood of becoming trapped in suboptimal solutions. For $\lambda=0$ the local minima present are the product states $\ket{00..}$ and $\ket{11..}$ and for $\lambda=1$ the local minima are defined by having two or more smaller GHZ states that are not entangled with each other.

This confirms the insight we obtained from studying the AKLT. It also confirms our suspicion that the choice of the loss function is an important factor in which local minima are present and how strongly they attract the optimizer. 

We note however that the difference between the use of our local minima avoidance minima and standard ADAM is not as large as the one seen for the AKLT. This might be due to the smaller system sizes that were used for this analysis, which are less prone to local minima, but are easier to perform statistics on.

\section{Optimizing the Energy}

In the main text, infidelity was employed as a metric to evaluate how accurately a given protocol approximates the AKLT state. In this appendix, the focus shifts to examining the effects of optimizing the energy instead of the infidelity. The Hamiltonian used for this purpose consists of two components:
\begin{equation}
H = M_{1\rightarrow 1/2} H^{1}_\text{AKLT} + \left[1 - M_{1\rightarrow 1/2} M_{1\rightarrow 1/2}^\dagger \right]
\end{equation}
The first term corresponds to the spin-1 AKLT Hamiltonian mapped to a spin-1/2 representation using the operator $M_{1\rightarrow 1/2}$. The second term is a projector onto the forbidden computational subspace. It enforces that the two spin-1/2 particles form a valid spin-1 state in the ground state.

\begin{figure*}[t]
    \centering
    \subfloat[]{\label{fig:mvqe:energyopt:ene}\includegraphics[width=0.45\textwidth]{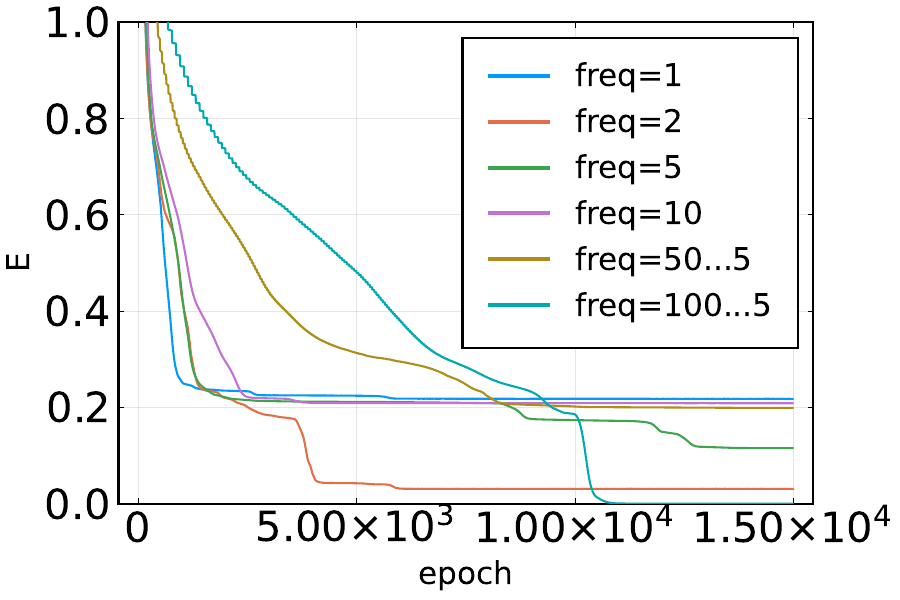}} 
    \subfloat[]{\label{fig:mvqe:energyopt:fid}\includegraphics[width=0.45\textwidth]{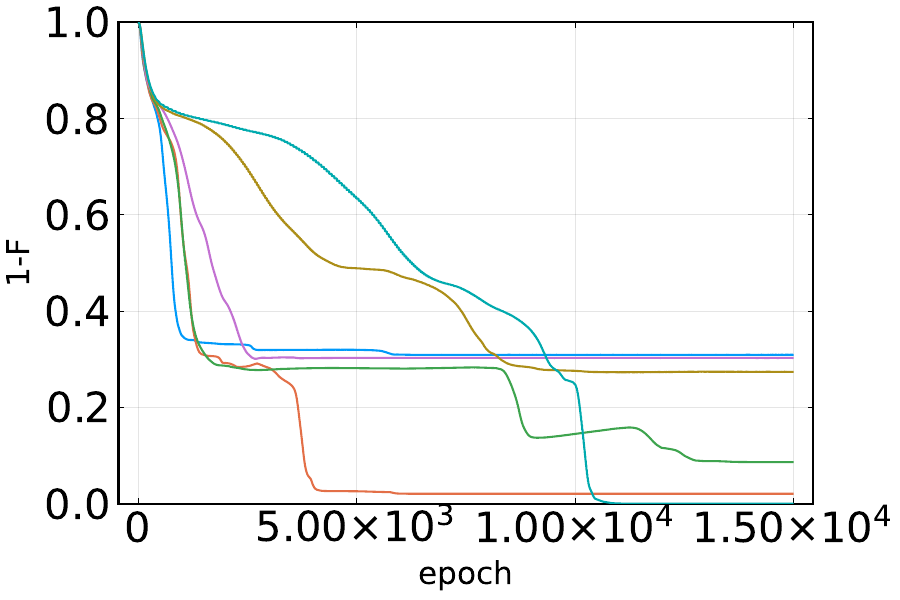}}
    
    \caption{Same optimization as in Fig.~\ref{fig:mvqe:freq_ancilla_regu=1}, but instead of using the infidelity as a cost function the energy is optimized. In panel (a) the energy is plotted against optimization epochs and in panel (b) the fidelity is computed for the same optimization runs.}
    \label{fig:mvqe:energyopt}
\end{figure*}

When energy is used as the objective function, the optimization landscape becomes significantly more complex. As shown in Fig.~\ref{fig:mvqe:energyopt}, only one optimization run successfully converged to a local minimum. By contrast, employing fidelity as the objective led to successful convergence in nearly all optimization runs, as illustrated in Fig.~\ref{fig:mvqe:freq_ancilla_regu=1}.

Using infidelity as the figure of merit results in a landscape where only the target state acts as an attractor. In contrast, energy-based optimization renders all eigenstates attractors. This increases the number of local minima and complicates the optimization process. Therefore, when fidelity is available, it is generally preferable to avoid using energy as the objective.

From a classical optimization perspective, fidelity also provides computational advantages. Its gradients are typically less costly to evaluate. The energy gradient takes the form  
\begin{align}
\bra{\frac{\partial\psi(\theta)}{\partial \theta}}H\ket{\psi} \,,
\end{align}  
whereas the gradient of the fidelity is given by  
\begin{align}
\braket{\psi_\text{target}|\psi(\theta)}\braket{\frac{\partial\psi(\theta)}{\partial \theta}|\psi_\text{target}} \,.
\end{align}  
In most cases, the target state $\ket{\psi_\text{target}}$ exhibits much lower entanglement than $H\ket{\psi}$. Consequently, fidelity gradients can be computed more efficiently than energy gradients when using Matrix Product States.

\section{Recurrent Neural Network}
\label{appendix:mvqe:rnn}

\begin{figure*}[ht]
\centering
\includegraphics[width=0.5\textwidth]{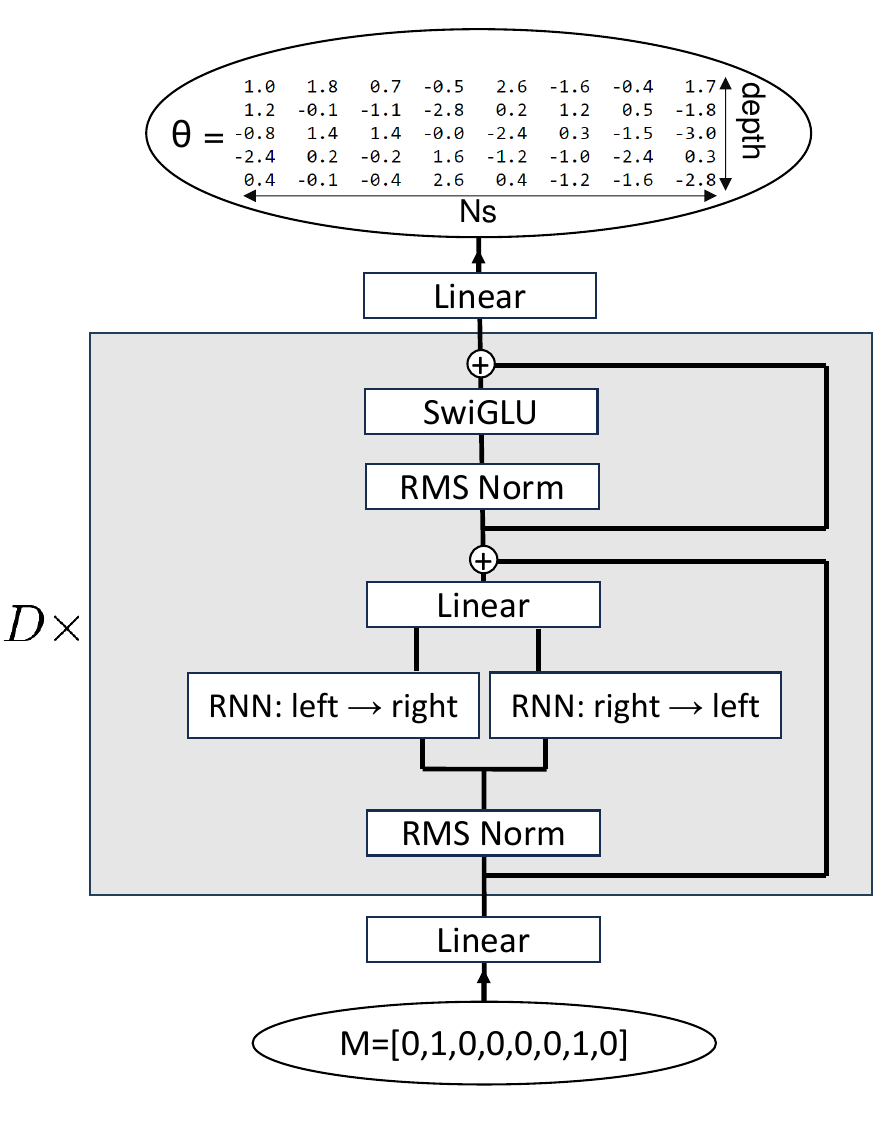}
\caption{\label{fig:mvqe:rnn:diagramm}
The architecture of the Recurrent Neural Network (RNN) used in this work. The input measurements pass through RMS normalization, a bidirectional RNN, and SwiGLU layers, generating the output angles $\theta$ for feedback unitaries. The grey box indicates repeated application of these core layers up to depth $D$.
}
\end{figure*}

The Recurrent Neural Network (RNN) structure used in this work is illustrated in Fig.~\ref{fig:mvqe:rnn:diagramm}. This architecture is derived from the Llama 3 model~\cite{dubey2024llama}, with the self-attention mechanism replaced by an RNN.

The RNN was selected for its ability to capture spatial dependencies necessary for inferring feedback gates from measurement outcomes. Both a bidirectional RNN as well as a unidirectional RNN were used. Note that when using a unidirectional RNN the first layer was switched from a Linear layer to a 1D Convolution with filter size 5, in order to make sure that the RNN could see enough measurement results to its right.

To mitigate the vanishing gradient problem often encountered in RNN training, a Gated Recurrent Unit (GRU)\cite{cho2014learning} was employed. GRUs address vanishing gradients by controlling how information is updated and forgotten, thus enabling effective learning over extended sequences. Long Short-Term Memory (LSTM) units were also considered, but they demonstrated inferior performance compared to GRUs.

Despite its strengths, the RNN architecture has limitations. Specifically, it struggles with long-range dependencies, reducing its effectiveness for large system sizes compared to transformer-based models. Transformer architectures, which rely on self-attention, may offer better gradient flow, particularly for long sequences. Future work could explore transformer or mamba-based models to potentially enhance performance.

\section{Feedback Unitaries}
\label{app:hardware_efficeint_two}

\begin{figure*}[t]
    \centering
    \subfloat[]{\label{fig:mvqe:feedback_unitaries:d_c=d}\includegraphics[width=0.45\textwidth]{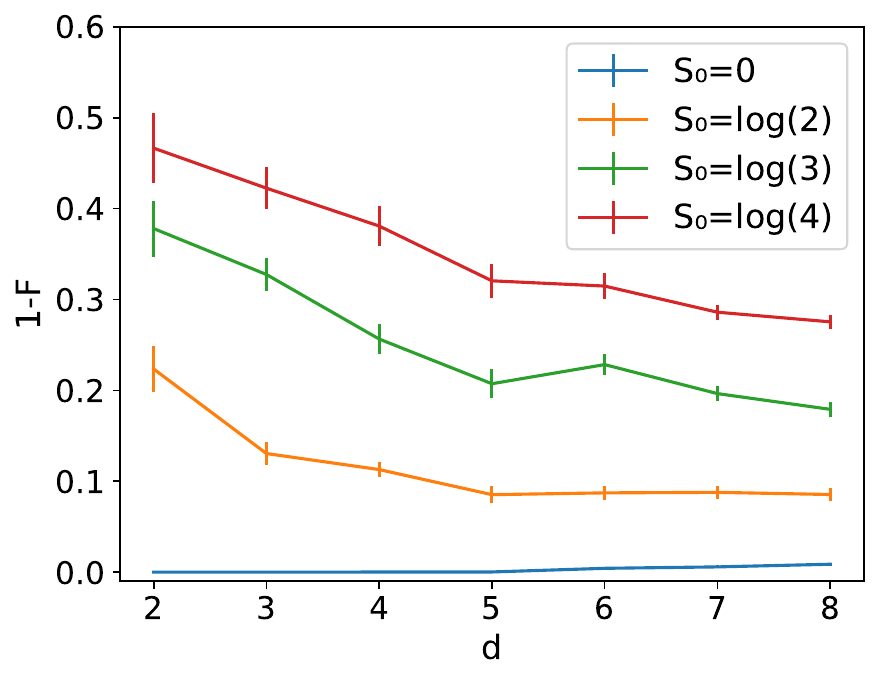}} 
    \subfloat[]{\label{fig:mvqe:feedback_unitaries:d_c=3}\includegraphics[width=0.6\textwidth]{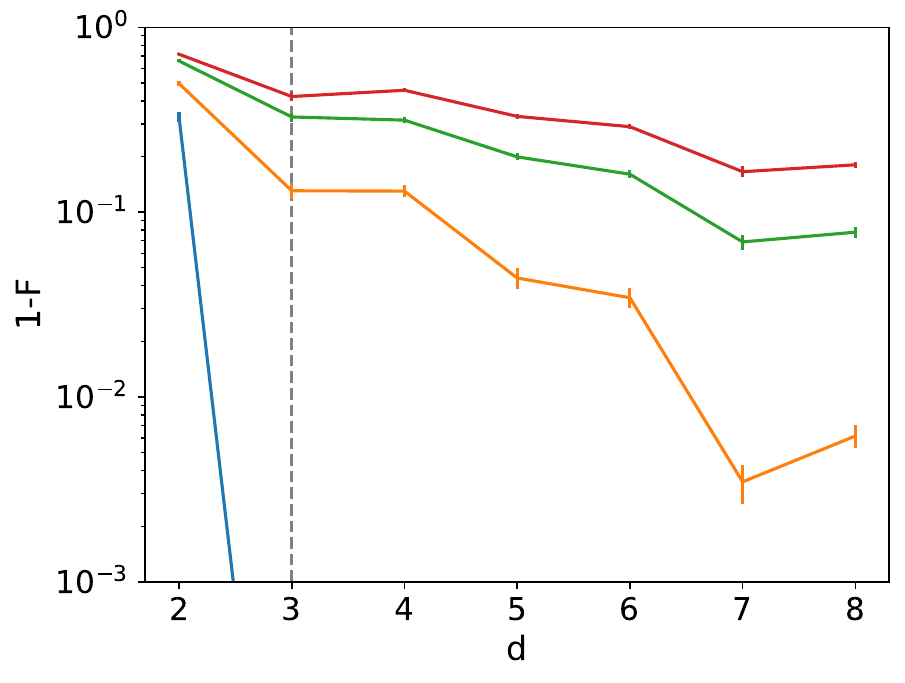}} \\
    \subfloat[]{\label{fig:mvqe:feedback_unitaries:circuit}\includegraphics[width=0.5\textwidth]{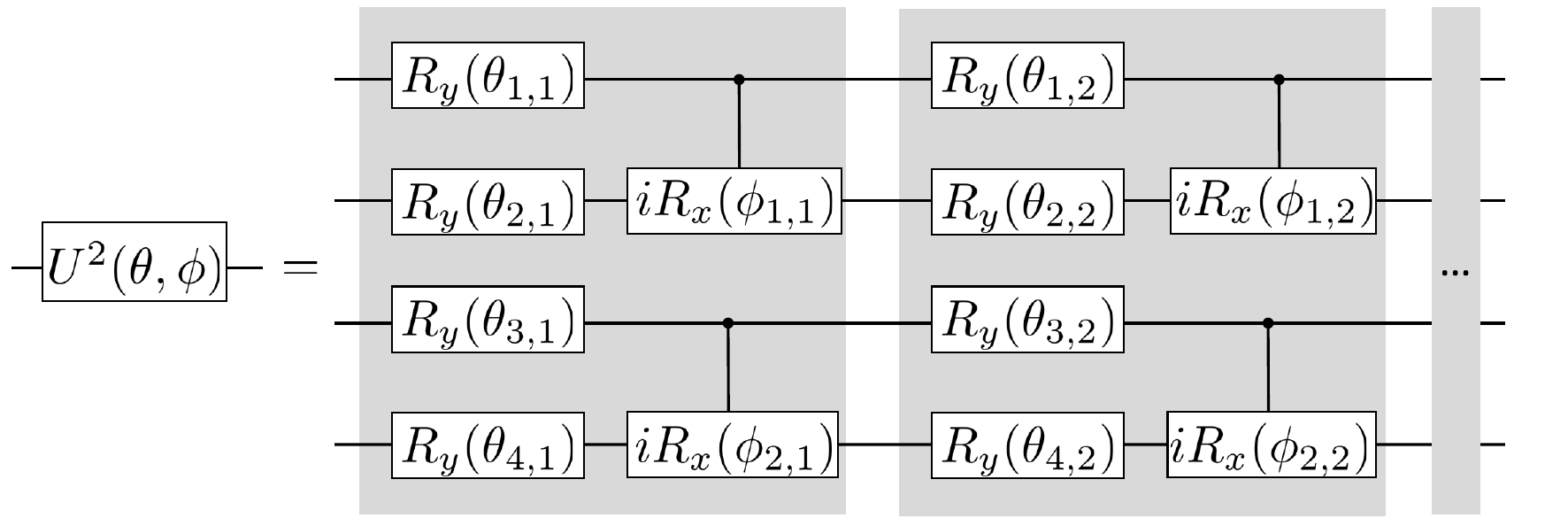}}
    
    \caption{(a) Plot showing the minimally obtained infidelity when trying to learn a target state prepared by a hardware-efficient ansatz as a function of circuit depth $d$ for different initial wave functions with entropies $S_0$. The teacher-student approach was used, where the teacher circuit was used to generate the target state, and the student circuit attempted to replicate it. (b) Infidelity as a function of circuit depth with the teacher circuit depth fixed at $d_t = 3$. The plot shows that infidelity decreases with increasing student circuit depth, but at a slow rate. (c) Diagram of the feedback ansatz used, showing the gate structure capable of representing any two-qubit gate at a depth of five.}
    \label{fig:mvqe:feedback_unitaries}
\end{figure*}%
Deciding on the optimal ansatz for the feedback step of the protocol is a nontrivial task. The goal is to use an ansatz that avoids creating long-range entanglement, ensuring that any observed long-range entanglement is entirely due to the measurement process.

To achieve this, a short hardware-efficient ansatz with a structure similar to that depicted in Fig.~\ref{fig:mvqe:ansatz} is ideal. However, short ansätze of this type are prone to numerous local minima~\cite{anschuetz2022swamped, cerezo2021shallow}. States that are close in Hilbert space may be significantly distant in parameter space, complicating optimization. This becomes specifically difficult when the initial state the variational circuit is applied on carries some entanglement. This was observed when trying to learn the feedback but is laid out here with the help of a simple toy problem.

This can be demonstrated by generating two random parameter sets, $\theta$ for the teacher and $\theta'$ for the student, and attempting to optimize $\max_{\theta'} \bra{\psi_0}U(\theta)^\dag U(\theta')\ket{\psi_0}$ for various depths. Although a minimum exists at $\theta' = \theta$, the ability of the optimization algorithm to locate it depends on the prevalence of local minima in the loss landscape.

Figure~\ref{fig:mvqe:feedback_unitaries:d_c=d} illustrates this calculation for a hardware-efficient ansatz, where the initial state $\ket{\psi_0}$ is a random MPS with bipartite entropy $S_0$ and a system size of eight qubits. When the initial entropy is $S_0 = 0$, the teacher-student infidelity remains close to zero across all circuit depths, indicating an absence of significant local minima in the loss landscape. However, for initial states with nonzero entropy, such as in the feedback protocol where $\ket{\psi_0}_S = \bra{M}_A U_1 \ket{0}_{S,A}$, the figure demonstrates that increasing circuit depth helps reduce the number of local minima, but the improvement is not substantial. The initial entropy of the quantum state has a significant impact on the complexity of the loss landscape.

To evaluate the difficulty of reproducing a quantum state prepared at a specific depth, the teacher circuit depth is fixed at $d_t = 3$. As shown in Fig.~\ref{fig:mvqe:feedback_unitaries:d_c=3}, for $S_0 \neq 0$ the infidelity decreases with increasing student circuit depth, but only slowly. These two plots provide insight into the complexity of the loss landscape for shallow circuits with entangled initial states, indicating that such circuits are not suitable if the initial state has some entanglement.

To optimize the circuit effectively, a greater circuit depth would be required, which is undesirable in this context. The aim is to demonstrate that entanglement is primarily propagated by the measurement process, rather than by the unitary gates. Therefore, to avoid creating excessive entanglement, an alternative approach was adopted. Instead of using a standard hardware-efficient ansatz, a sparsely connected ansatz was employed (see Fig.~\ref{fig:mvqe:feedback_unitaries}). This ansatz utilizes a two-body gate:

\begin{align}
CiRX(\theta) = 
\begin{pmatrix}
1 & 0 & 0 & 0 \\
0 & 1 & 0 & 0 \\
0 & 0 & \cos\left(\frac{\theta}{2}\right) & \sin\left(\frac{\theta}{2}\right) \\
0 & 0 & \sin\left(\frac{\theta}{2}\right) & \cos\left(\frac{\theta}{2}\right),
\end{pmatrix}
\end{align}
that can interpolate between the CNOT and the identity gate. The ansatz is capable of representing any two-qubit gate at a depth of five, which was used in all experiments. Notably, the teacher-student infidelity for a depth of five is indeed zero.

\section{Left correctability of the Protocol}
\label{app:mvqe:leftcorrectable}

In this section, the method used to determine if a protocol's feedback can be corrected with information from only one direction is presented. This is relevant as for example in the protocol of Smith~\etal~\cite{smith2023deterministic} only information from one side is necessary to establish the feedback gates. Using the technique described here, it was confirmed that all protocols developed in this work are indeed left correctable.

The feedback angles are first determined for a specific measurement outcome $M$. Then, the stability of these angles is assessed by modifying individual bits in the measurement outcome and observing the resulting changes in the feedback angles.

\begin{figure*}[hbt]
\centering
\includegraphics[width=0.5\textwidth]{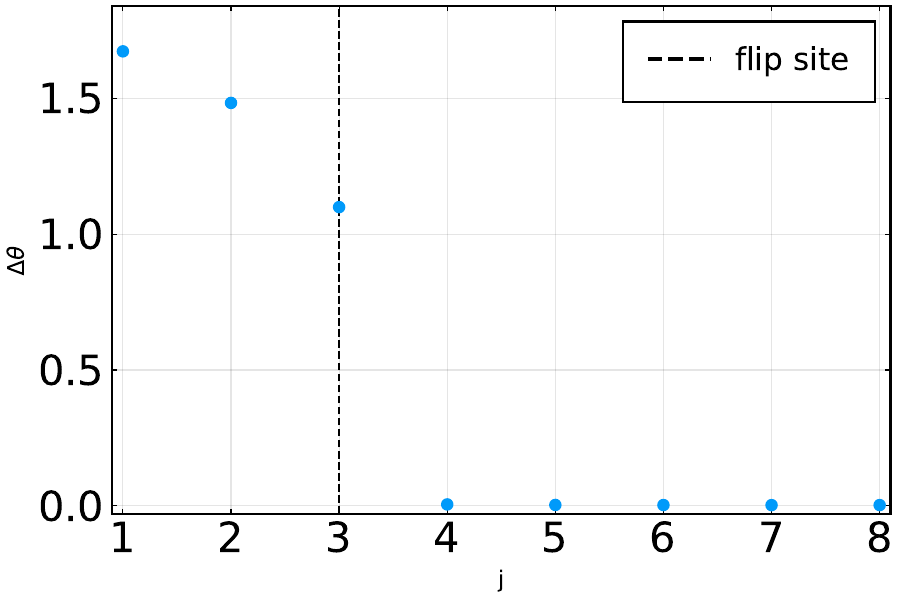}
\caption{
Figure showing the difference in feedback angles, $\Delta\theta$, for each site $j$ after flipping a specific bit in the measurement outcome. The black vertical line represents the site where the bit was flipped.
}
\label{fig:mvqe:left_correctable}
\end{figure*}

The algorithm proceeds as follows: given a measurement outcome $M$, the optimal feedback angles $\theta$ are determined. Subsequently, $M$ is modified by flipping a specific bit, indicated by `index`, resulting in a new measurement outcome $M'$. The new set of feedback angles $\theta'$ is then determined for $M'$. The difference $\Delta\theta$ between $\theta$ and $\theta'$ serves as an indicator of whether the feedback mechanism can be corrected using information from a single direction. As an example this procedure was done for the AKLT and plotted in Fig.~\ref{fig:mvqe:left_correctable}.

\begin{algorithm}[H]
\caption{Feedback Correctability Check}
\label{alg:mvqe:left_corectable}
\begin{algorithmic}
\Function {left\_correctable}{index}
    \State $\ket{\psi_1} \gets U_1\ket{0}$
    \State $M \gets \text{sample}_A(\ket{\psi_1})$
    \State $\theta \gets \max_\theta\left[\bra{\psi_\text{target}} U_2(\theta)\bra{M}\ket{\psi_1}\right]$
    \State $M' \gets \text{flipbit}(M, \text{index})$
    \State $\theta' \gets \max_\theta\left[\bra{\psi_\text{target}} U_2(\theta)\bra{M'}\ket{\psi_1} - \lambda \sum_{i, j=\text{index}}(\theta_{i,j}- \theta'_{i,j})^2  \right]$
    \State $\Delta\theta = |\theta - \theta'|$
    \State \Return $\Delta\theta$
\EndFunction
\end{algorithmic}
\end{algorithm}

\end{document}